\documentclass[a4paper,fleqn]{cas-dc}

\usepackage[numbers,sort&compress]{natbib}

\usepackage{amsmath}
\usepackage{booktabs}
\usepackage{multirow}
\usepackage{siunitx}
\usepackage{tikz}
\usepackage{graphicx}
\usepackage{amsmath, amssymb}
\usepackage{dirtytalk}
\usepackage[section]{placeins}

\DeclareUnicodeCharacter{2212}{\textminus}

\begin{document}
\let\WriteBookmarks\relax
\def\floatpagepagefraction{1}
\def\textpagefraction{.001}
\usetikzlibrary{shapes.geometric, arrows.meta, positioning}

\tikzstyle{startstop} = [rectangle, rounded corners, minimum width=3cm, minimum height=1.5cm,
                         text centered, draw=black, fill=gray!15, text width=8cm, align=center]
\tikzstyle{process}   = [rectangle, minimum width=3cm, minimum height=1.5cm,
                         text centered, draw=black, fill=blue!10, text width=8cm, align=center]
\tikzstyle{arrow}     = [thick, ->, >=stealth]

\shorttitle{Magnetic Contributions to Phase Stability in the Co--Ni Binary}
\shortauthors{P. Jalagam et al.}

\title[mode=title]{Magnetic Contributions to Phase Stability in the Co--Ni Binary: A First-Principles CALPHAD Study}

\author[1,2]{Prajna Jalagam}[orcid=0009-0008-7289-904X]
\credit{Conceptualization, Methodology, Validation, Formal analysis, Investigation, Data curation, Visualization, Writing -- original draft, Writing -- review \& editing}

\author[2]{Zhigang Wu}[orcid=0000-0001-8959-2345]
\credit{Validation, Supervision, Writing -- review \& editing}

\author[2]{John W. Lawson}[orcid=0000-0002-1504-4829]
\credit{Resources, Supervision, Funding acquisition, Writing -- review \& editing}
\cormark[1]
\ead{john.w.lawson@nasa.gov}

\author[1]{Axel {van de Walle}}[orcid=0000-0002-3415-1494]
\credit{Conceptualization, Methodology, Validation, Supervision, Funding acquisition, Writing -- review \& editing}
\cormark[1]
\ead{avdw@brown.edu}

\affiliation[1]{organization={School of Engineering, Brown University},
                addressline={Box D},
                city={Providence},
                state={RI},
                postcode={02912},
                country={USA}}

\affiliation[2]{organization={Intelligent Systems Division, NASA Ames Research Center},
                city={Moffett Field},
                state={CA},
                postcode={94035},
                country={USA}}

\cortext[1]{Corresponding author}

\begin{abstract}
We propose a simple method to employ ab-initio calculations to determine magnetic contributions to free energy of alloys. Validation on the Co–Ni binary demonstrates that this ab initio approach reproduces experimental FCC–HCP phase equilibria while providing physically transparent, structure-dependent magnetic parameters suitable for multicomponent extrapolation. Our results demonstrate that physically grounded magnetic parametrization, derived directly from electronic structure calculations, enables predictive phase diagram modeling for magnetic alloy systems.
\end{abstract}


\begin{keywords}
Magnetic Gibbs energy \sep Disordered local moment \sep Curie temperature \sep Special quasirandom structures \sep Co-Ni
\end{keywords}

\maketitle

\section{Introduction to Co-Ni binary alloy CALPHAD studies}
\label{sec:intro}

\begin{figure}[pos=!htbp]
    \centering
    \includegraphics[width=1\linewidth]{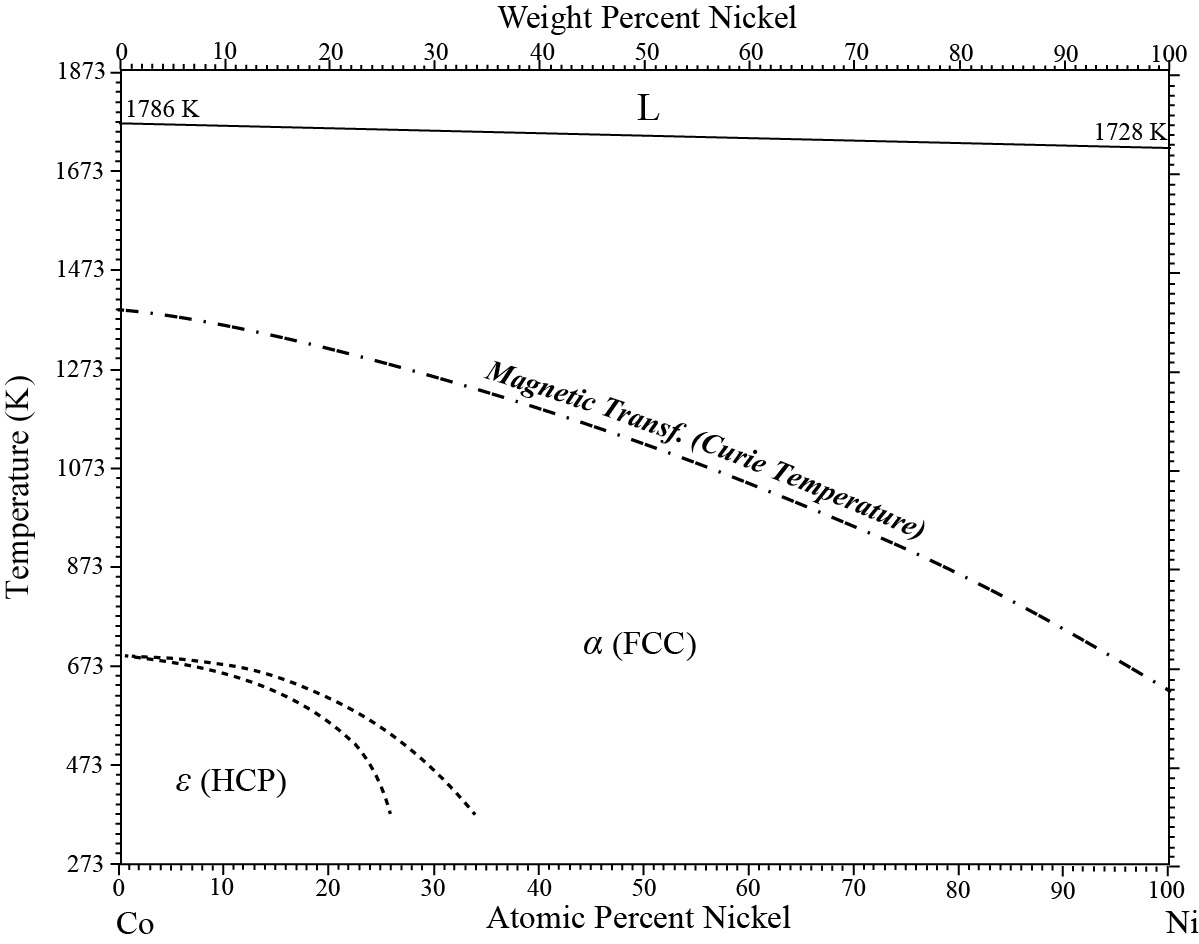}
    \caption{Previous experimental assessment of magnetic transition temperature and phase stability of the Co-Ni binary~\cite{FernandezGuillermet}}
    \label{fig:enter-label}
\end{figure}

The Co–Ni binary alloy system plays a central role in the development of high-performance structural and functional materials, including superalloys for aerospace applications, magnetic alloys for energy conversion, and foundational base systems for high-entropy alloys. The system has been extensively characterized, summarized as shown in Figure~\ref{fig:enter-label} by a mostly isomorphous phase diagram exhibiting by solid solubility in both face-centered cubic (FCC, $\alpha$) and hexagonal close-packed (HCP, $\varepsilon$) lattices, along with robust ferromagnetism across the entire composition range~\cite{nishizawa1983co_ni, Asadikiya2020, Kaufman1978, Liu2005, Sun2015}. These features, combined with the absence of stable intermetallic compounds, make Co–Ni an ideal platform for investigating thermodynamic modeling strategies that integrate first-principles calculations with CALPHAD (CALculation of PHAse Diagrams) methodologies, particularly in systems where magnetism plays a critical role in phase stability. \smallskip

While the Guillermet 1987~\cite{FernandezGuillermet} assessment of the Co-Ni binary outlined above is widely accepted among the CALPHAD community, more recent efforts~\cite{SHI2025121476} bring into question the FCC $\leftrightarrow$ HCP transition phase boundaries by reassessing vibrational and magnetic contributions to the free energy using first-principles. While this work assesses vibrational entropy to be the main contributing factor to shifting phase boundaries, significant gaps remain in the thermodynamic modeling of magnetic contributions to phase stability, particularly in the case of metastable phases such as HCP-Ni. The magnetic moment ($\beta$) and Curie temperature ($T_{C}$) for HCP-Ni are not experimentally measurable, and are missing in standard thermodynamic databases such as SGTE~\cite{sgte1991}. Widely accepted databases~\cite{LIU2016125} repurposed the magnetic parameters and non-ideal magnetic behavior in solid solutions of FCC-Ni for HCP-Ni (these parameters are listed in Table~\ref{tab:tc_bm}, the physical significance of which is outlined in Sections~\ref{sec:calphadfreee} and~\ref{sec:magneticmodels}), while newer works~\cite{SHI2025121476} present first-principles calculated values that deviate. Overall, accurate extrapolation of these quantities is essential for constructing consistent CALPHAD models that include metastable phase extensions across the full composition and temperature range. The associated thermodynamic consequences of temperature-dependent magnetic entropy on phase boundaries for the Co-Ni binary remain underexplored. \smallskip

\begin{table}[pos=!htbp]
    \centering
    \caption{CALPHAD magnetic model parameters for the Co-Ni binary~\cite{FernandezGuillermet}}
    \label{tab:tc_bm}
    \setlength{\tabcolsep}{3pt}
    \begin{tabular}{l
                    S[table-format=4.0] S[table-format=3.0]
                    S[table-format=3.0] S[table-format=4.0]
                    S[table-format=1.2]  S[table-format=1.2]
                    S[table-format=1.3]  S[table-format=1.3]}
        \toprule
        & \multicolumn{4}{c}{$T_C$ (K)}
        & \multicolumn{4}{c}{$\beta$ ($\mu_\text{B}$)} \\
        \cmidrule(lr){2-5} \cmidrule(lr){6-9}
        Phase & {Co} & {Ni} & {$L_0$} & {$L_1$}
              & {Co} & {Ni} & {$L_0$} & {$L_1$} \\
        \midrule
        FCC\_A1 & 1396 & 633 & 411 & -99 & 1.35 & 0.52 & 1.046 & 0.165 \\
        HCP\_A3 & 1396 & 633 & 411 & -99 & 1.35 & 0.52 & 1.046 & 0.165 \\
        \bottomrule
    \end{tabular}
\end{table}

This paper introduces a framework based on Special Quasirandom Structures (SQS) to determine magnetic parameters for established semi-empirical CALPHAD models, allowing for more accurate free energy descriptions of stable and metastable phases of materials with magnetic character, applied to the Co-Ni solid solution binary system. \smallskip

\section{CALPHAD Free Energy Descriptions for Phases}
\label{sec:calphadfreee}
In CALPHAD modeling, the molar Gibbs energy of a phase $\theta$ is
expressed as~\cite{lukas2007computational}
\begin{equation}
    G^{\theta} = G_{\mathrm{srf}}^{\theta}
    + G_{\mathrm{phys}}^{\theta}
    + G_{\mathrm{cnf}}^{\theta}
    + G_{\mathrm{xs}}^{\theta}
    \label{eq:gibbs}
\end{equation}
where all quantities are molar, the superscript $\theta$ labels the phase, and
each contribution is defined as follows.
\begin{itemize}
    \item $G_{\mathrm{srf}}^{\theta}$ is the surface of reference, the Gibbs
    energy of an unreacted mixture of the constituents of the phase,
    $G_{\mathrm{srf}}^{\theta} = \sum_i x_i\, {}^{0}G_i^{\theta}(T)$, where
    ${}^{0}G_i^{\theta}(T)$ is the temperature-dependent molar Gibbs energy of pure
    component $i$ in structure $\theta$. The vibrational contribution is obtained from Born--von Karman phonon calculations with ATAT's \texttt{fitfc} and incorporated through the \texttt{sqs2tdb} solution-model fit. The excess vibrational entropy of mixing is found to be negligible across the binary (see Section~\ref{sec:firstprinciple}).
    \item $G_{\mathrm{phys}}^{\theta}$ denotes the Gibbs energy of a physical
    model. Here we consider the magnetic model, so $G_{\mathrm{phys}}^{\theta} =
    G_{\mathrm{mag}}^{\theta}$. The CALPHAD magnetic models used for
    $G_{\mathrm{mag}}^{\theta}$ are outlined in Section~\ref{sec:magneticmodels}.
    The focus of this work is to verify and obtain the data that parametrizes
    this contribution.
    \item $G_{\mathrm{cnf}}^{\theta}$ denotes the Gibbs energy from
    configurational entropy, $G_{\mathrm{cnf}}^{\theta} = -T\,S_{\mathrm{cnf}}^{
    \theta}$, based on the number of arrangements of the constituents. For the
    disordered solid solutions considered here it reduces to random mixing on
    each sublattice.
    \item $G_{\mathrm{xs}}^{\theta}$ is the excess Gibbs energy, the remaining
    non-ideal part of the real Gibbs energy once the first three terms are subtracted. It
    carries no separate physical model and absorbs the residual vibrational,
    electronic, and other contributions not already captured.
\end{itemize}
The compositional dependence of physical model parameters or the thermodynamic quantities themselves is commonly expressed by combining a linear endmember reference with a Redlich--Kister (RK) polynomial to capture non-ideal mixing. For a binary system A-B, a general property $P(x)$ is expressed as
\begin{equation}
    P(x) = x_A P_A + x_B P_B + x_A x_B \sum_{n=0}^{N} L_n^{(P)}(x_A - x_B)^n
    \label{eq:rk}
\end{equation}
where $x_A$ and $x_B$ are the mole fractions of components A and B. The values $P_A$ and $P_B$ are the corresponding endmember parameter values, forming the linear reference. The coefficients $L_n^{(P)}$ are the Redlich--Kister interaction parameters, which apply when the property deviates from ideal linearly interpolated behavior.\smallskip 

When applying this formalism to the molar Gibbs energy of Eq.~\eqref{eq:gibbs}, the linear reference terms correspond exactly to the surface of reference, $G_{\mathrm{srf}}^{\theta}$, while the Redlich--Kister summation represents the excess Gibbs energy, $G_{\mathrm{xs}}^{\theta}$. For physical model parameters like $T_C(x)$ and $\beta(x)$, Eq.~\eqref{eq:rk} is applied in its entirety to interpolate their values across the composition space. The configurational term $G_{\mathrm{cnf}}^{\theta}$ is derived from statistical thermodynamics and does not utilize this expansion.\smallskip

For the Gibbs energy, the interaction parameters are temperature dependent, commonly fit in the form $L_n^{(G)} = a + bT + cT\ln{T}$ via DFT-obtained phonon calculations and formation energies. For the magnetic properties, by contrast, the Redlich--Kister coefficients are temperature-independent numerical constants. $T_C(x)$ and $\beta(x)$ are each represented by Eq.~\eqref{eq:rk} with $P = T_C$ or $P = \beta$, giving a pure polynomial in composition whose endmembers and interaction coefficients carry units of $\mathrm{K}$ and $\mu_B/\mathrm{atom}$ respectively and have no temperature dependence. These composition-dependent quantities then enter the CALPHAD magnetic models, where $T_C$ sets the reduced temperature $\tau = T/T_C$ and $\beta$ defines the magnetic entropy limit (further outlined in Section~\ref{sec:magneticmodels}).\smallskip

\section{CALPHAD Magnetic Models} \label{sec:magneticmodels}
Classical CALPHAD treatments of magnetism in Co–Ni alloys rely on variations of the Inden–Hillert–Jarl (IHJ) model. Inden’s original model~\cite{Inden1976} introduced a closed-form expression for the magnetic contribution to heat capacity and as a result entropy, using a mean magnetic moment ($\beta$) per atom. Hillert and Jarl~\cite{Hillert1978} reformulated Inden's model by replacing the logarithmic terms with a power series expansion in `reduced temperature' so that it could be implemented as a standard Gibbs energy contribution in CALPHAD software. In the IHJ model, the magnetic Gibbs energy is expressed as: 
\begin{equation}
    G_{mag}(T) =-RT\ln(\beta + 1)f(\tau)
    \label{eq:Gmag}
\end{equation}
where $\tau=T/T_\text{C}$, $T_\text{C}$ is Curie temperature, and $f(\tau)$ is a piecewise function that is a universal system-independent, structure-dependent dimensionless function that captures the long-range ordered (ferromagnetic) and paramagnetic states (see Appendix~\ref{sec:appendixA}).\smallskip

Composition dependence is supplied by Redlich-Kister polynomials fitted to experiment, and a non-magnetic constituent is handled by assigning the element a vanishing critical temperature and a vanishing moment so that the whole composition dependence falls to the binary interaction parameters. Where constituents differ in magnetic character, a single curve is made to span both regions by means of an antiferromagnetic factor~\cite{TauerWeiss1955,WeissTauer1956,HERTZMAN198267}. This scheme is used in the Co-Cr-Ni assessments~\cite{LIU2016125} that are the natural target of the present work, and Section~\ref{sec:magmom} returns to it.\smallskip

This IHJ model has been the standard in CALPHAD and is adopted in major thermodynamic databases. For Co–Ni solid solutions (both fcc $\alpha$ and hcp $\varepsilon$ phases), the IHJ model allows the ferromagnetic ordering contribution to Gibbs energy to be included as a function of composition via composition-dependent $T_\text{C}(x)$ and $\beta(x)$ fits, providing a semi-empirical magnetic Gibbs energy term. All conventional Co–Ni CALPHAD assessments to date have used these models (embedded in e.g. SGTE unary descriptions and solution phase models) as the basis for magnetic contributions, providing context for the parameters listed in Table~\ref{tab:tc_bm}.\smallskip

Successive revisions have addressed known deficiencies of this form~\cite{Chuang1985, vanBohemen2025}. We retain the IHJ expression throughout this work, with the corresponding dimensionless function for the revised model of Xiong et al.~\cite{Xiong2012}, referred to here as the Inden-Hillert-Xiong (IHX) model, given alongside $f(\tau)$ in Appendix~\ref{sec:appendixA}. Independently of that choice, Ref.~\cite{Xiong2012} draws a distinction in how $\beta$ itself is constructed that we make use of below: the mean moment reproduces the maximum magnetic entropy only approximately when the constituent atoms carry different local moments, while a geometric average (or ``effective") magnetic moment $\beta^{*}$ can be defined that is more consistent with the definition of entropy. Since $\beta$ enters Eq.\ref{eq:Gmag} only as an input, $\beta^{*}$ may be used with the IHJ form unchanged. We compute $\beta^{*}$ for each SQS from the species-resolved local moments
and write it simply as $\beta$ hereafter, i.e., the mean moment is not used in this
work.\smallskip

A fully statistical-mechanical alternative to this class of model is provided by Zentropy theory~\cite{Liu2022,Liu2024}, in which the free energy of a phase is obtained from a partition function over DFT-computed configurations, without relying on a shape function or structure factor. That route is the more fundamental one, and in principle avoids the semi-empirical nature of the IHJ form. Its application to the present problem, however, requires that the configurations entering the partition function represent all possible spin states of each atom. Since $\beta = 2S$, a local moment admits $\beta + 1$ orientational states, and it is precisely these that the IHJ entropy $R\ln(\beta + 1)$ counts, while collinear DFT only supplies two per site. Constructing a tractable set of spin configurations that spans all $\beta + 1$ states and weighting them correctly in the partition function could be a fruitful avenue for future research. We therefore retain the IHJ form throughout, both because its parameters are the quantities existing CALPHAD databases parametrize directly and because the first-principles input it requires, $T_{\rm C}$ and $\beta$ per phase, is obtainable at a cost that scales to the multicomponent systems of interest.\smallskip

\section{Methods} \label{sec:style}
\subsection{Special Quasirandom Structure (SQS) Generation and First Principles Calculations}
\label{sec:firstprinciple}
To model the disordered FCC and HCP solid solutions in the Co–Ni binary system, we use the sqs2tdb tools of the Alloy Theoretic Automated Toolkit (ATAT)~\cite{vandeWalle2017, vandeWalle2002}. Structural input files for total energy calculations were generated using the default SQS library provided by ATAT, yielding 32-atom supercells for most compositions and 64-atom supercells for HCP structures at 12.5, 37.5, 62.5, and 87.5 at.\% Co. For magnetic parameter extraction, we constructed separate 16-atom SQS cells at compositional increments corresponding to \texttt{lev=2} through \texttt{lev=5} (12.5–25 at.\% spacing) spanning pure Ni (0 at.\% Co) to pure Co (100 at.\% Co).\smallskip

First-principles calculations for all SQS configurations were computed using the Vienna Ab initio Simulation Package (VASP)~\cite{Kresse1996CMS, Kresse1996PRB} with the Perdew–Burke–Ernzerhof (PBE) exchange-correlation functional under the generalized gradient approximation (GGA)~\cite{Perdew1996}. Collinear spin-polarized calculations were used throughout, with an initial magnetic moment of +2 $\mu_{B}$ assigned to each atom to ensure convergence to the ferromagnetic ground state. The plane-wave kinetic energy cutoff (ENCUT) of 475 eV was used in the projector augmented wave (PAW) method~\cite{Blochl1994}. The k-point mesh was automatically generated by ATAT such that the total number of k-points per reciprocal atom was at least 8000. All SQS structures were fully relaxed with respect to ionic positions and cell shape as no endmembers and SQS were found to be mechanically unstable (defined as exhibiting a cell shape relaxation of over 5\%~\cite{vandeWalle2015, vandeWalle2017epicycle}) .\smallskip

The total energies obtained from DFT calculations for the default SQS configurations were used to derive formation enthalpies relative to ab initio-computed pure FCC and HCP Co and Ni end members. These formation enthalpies serve as the first-principles reference for mixing and excess energy calculations. To connect these results to experimentally assessed data, the free energies of pure Co and Ni were additionally sourced from the SGTE elemental database~\cite{sgte1991}. This allows the incorporation of empirical thermodynamic data for phase stability evaluations. The ab initio formation enthalpies were integrated into a thermodynamic framework using the \texttt{sqs2tdb -fit} utility within ATAT~\cite{vandeWalle2017}, and the resulting parameters were then embedded into a CALPHAD-compatible thermodynamic database via \texttt{sqs2tdb -tdb}~\cite{vandeWalle2017}. This enabled phase diagram construction and the quantitative assessment of phase stability.\smallskip

Vibrational free energy contributions were evaluated for endmembers to account for temperature-dependent effects on phase stability. Phonon calculations were performed using the Born–von Karman model~\cite{vandeWalle2002Phase}, with force constants derived from finite-displacement methods implemented in ATAT’s \texttt{fitfc} code~\cite{vandeWalle2009}. The interaction range for force constants was set to twice the nearest-neighbor distance ($\sim 5\,\text{\AA}$) to capture the dominant interatomic interactions. To verify that linear interpolation of the vibrational entropy between endmembers is sufficient, we performed phonon calculations on intermediate SQS compositions, including $X_{\mathrm{Co}} = 0.5$, for both FCC and HCP structures. While both phases exhibit a nonzero high-temperature limit of the vibrational entropy of formation, they demonstrate a strikingly similar composition-dependence. At $X_{\mathrm{Co}} = 0.5$, the FCC and HCP phases deviate from linearity by $-0.052 k_B$ and $-0.049 k_B$ per atom, respectively. Furthermore, the maximum difference in formation entropy between the two phases across all evaluated compositions is only $0.050 k_B$ per atom. Because both phases deviate in the same direction with nearly equivalent magnitudes, the excess vibrational entropy of mixing effectively cancels out. Thus, endmember-only calculations are adequate for the Co-Ni FCC $\rightleftharpoons$ HCP phase diagram calculation.

\subsection{Paramagnetism and Magnetic Transition Temperatures}\label{sec:paramag}\smallskip
\subsubsection{Estimating the Paramagnetic State: the Disordered-Local Moment (DLM) Approach}\label{sec:DLM}
Modeling the paramagnetic state at elevated temperatures poses a significant challenge for first-principles methods. In the paramagnetic regime (above the Curie temperature $T_C$ or Néel temperature $T_N$), long-range magnetic order is lost, yet local magnetic moments often persist and fluctuate in direction~\cite{Antropov2004, Ruban2007}. For example, pure Ni retains local moments above its $T_C\approx 631$ K, with strong short-range magnetic order among Ni atoms even in the disordered phase~\cite{Antropov2004}. Accurate simulation of such states is critical for predicting high-temperature properties (e.g. phase stability, thermodynamics, transport), meaning standard density functional theory (DFT) at
$T=0$ requires careful statistical mechanics to represent a disordered local moment ensemble. The Co–Ni binary alloy system
exemplifies the need for this approach: both Co and Ni are ferromagnetic at low temperature, yet in certain compositions or when mixed with other elements, the magnetic transition temperature can be drastically reduced. For instance, equiatomic Cr–Co–Ni has an experimentally reported Curie temperature of only ~4–5 K~\cite{APL_CrCoNi}. This underscores the need for robust methods and formulations to simulate paramagnetic states with DFT even in alloys containing strongly magnetic elements like Co and Ni.\smallskip

The Disordered Local Moment (DLM) picture provides a framework to include thermally induced spin fluctuations in a first-principles description. In the DLM approach, one assumes that well-defined local magnetic moments exist on atomic sites, but their orientations are randomly disordered (paramagnetic) rather than aligned. Importantly, these local moments are treated as long-lived with respect to electronic motions, analogous to a Born–Oppenheimer separation between the spin configuration and the electrons~\cite{Kudrnovsky2012, Kudrnovsky2013} This allows one to solve the electronic structure for a fixed representative configuration of disordered spins, then statistically average over many such configurations to mimic the paramagnetic state. Early implementations of DLM were formulated in combination with the coherent potential approximation (CPA) in KKR or LMTO electronic structure methods. There, the paramagnetic state is treated like a fictitious alloy of “spin-up” and “spin-down” moments on each site, averaged via CPA. This formalism was pioneered by Gyorffy, Staunton, Stocks and co-workers in the 1980s~\cite{Gyorffy1985}, enabling calculations of paramagnetic Fe, Co, Ni and their alloys.\smallskip

Recent implementations of the disordered local moment (DLM) model in plane-wave DFT codes, such as VASP, have enabled supercell-based simulations of paramagnetic states by randomly assigning spin-up and spin-down moments to magnetic atoms within the cell to ensure zero net magnetization~\cite{Koermann2012, Ruban2007, Ma2017, Su2024} and uncorrelated spin orientations. In systems like Co--Ni, this approach is realized by constructing special quasirandom structures (SQSs) with equal numbers of Co ${\uparrow}$, Co ${\downarrow}$, Ni ${\uparrow}$, and Ni ${\downarrow}$ atoms, effectively mimicking an Ising-like disordered moment ensemble~\cite{Koermann2012, Ma2017, Su2024}. These spin-disordered supercells preserve local magnetic moments and avoid the limitations of spin-unpolarized (nonmagnetic) calculations, making them suitable for modeling the paramagnetic phase~\cite{Kudrnovsky2012}. \smallskip

We extend the special quasirandom structure (SQS) formalism~\cite{Zunger1990} to include spin disorder in addition to compositional disorder. The traditional SQS approach captures chemical randomness by arranging multiple species to replicate the target pair and multisite correlation functions of a random alloy~\cite{Zunger1990}. Analogously, spin-up and spin-down moments can be treated as pseudo-species, yielding a collinear spin-SQS whose low-order spin-pair correlations vanish as in an idealm paramagnet~\cite{Drautz2004}. This construction minimizes artificial magnetic ordering in small supercells, a common artifact of naive random spin assignments, and captures the essential features of paramagnetic disorder in a single representative structure without configurational averaging over many disordered magnetic states~\cite{Drautz2004, Su2024}. Relative to single-site DLM treatments, the explicit supercell additionally allows each local moment to respond to its chemical and magnetic environment, retaining near-neighbor effects that a mean-field average suppresses~\cite{Ruban2007}. The $T = 0$ spin-SQS total energies characterize the energetics of the disordered state; the associated magnetic entropy is incorporated at the thermodynamic-modeling stage, as in established DLM-based treatments of finite-temperature magnetism~\cite{Ruban2007, Ma2017}, here through the frameworks described later in Sec.~\ref{sec:curie-weiss}.

The use of DLM with the SQS formalism has led to improved predictions of high-temperature material properties. For instance, paramagnetic phonon spectra calculated for bcc and fcc iron show excellent agreement with experiment, capturing the dynamic stabilization of high-$T$ phases that are unstable in conventional FM or NM calculations~\cite{Yang2023, Koermann2012}. Recent work by Su \textit{et al.}~\cite{Su2024} further validates this approach by demonstrating that spin-chemical cluster expansions parameterized by ab-initio calculated energies from DLM-configured spin disorder on compositional SQS supercells enable accurate predictions of short-range order and phase stability in Fe–Ni–Cr alloys. By employing SQS structures as unbiased disordered reference states, this framework allows physically meaningful chemical and magnetic short-range order to emerge naturally through statistical mechanical sampling, without artifacts arising from artificially ordered reference configurations. Their spin cluster expansion framework captures the coupled effects of magnetic and chemical disorder through statistical mechanical sampling, outperforming conventional models based on magnetically ordered reference states, particularly in the high-temperature paramagnetic regime.\smallskip

\begin{figure}[pos=!htbp]
\centering
\resizebox{\linewidth}{!}{
\begin{tikzpicture}[node distance=1.9cm]
\node (start) [startstop] {\textbf{Start:} Obtain \texttt{clusters.out} and \texttt{sqscell.out} from \texttt{sqs2tdb} for single sublattice structure};
\node (compSQS) [process, below of=start] {Generate compositional SQS using \texttt{sqs2tdb} or \texttt{mcsqs} at chosen compositional coarseness (\texttt{lev})};
\node (spinSQS) [process, below of=compSQS] {Generate 50:50 spin SQS (↑/↓) using \texttt{mcsqs}, preserving cell shape (\texttt{sqscell.out}) and atom count (-n)};
\node (randomspin) [process, below of=spinSQS] {Map spin SQS to compositional SQS using custom \texttt{randomspin} code with constraint $\sum_i$ \texttt{MAGMOM}$ = 0$};
\node (runvasp) [process, below of=randomspin] {Run collinear spin-polarized DFT relaxations using \texttt{runstruct\_vasp} or \texttt{robustrelax\_vasp}};
\node (end) [startstop, below of=runvasp] {\textbf{Output:} Energy estimate for the DLM paramagnetic approximation state};

\draw [arrow] (start) -- (compSQS);
\draw [arrow] (compSQS) -- (spinSQS);
\draw [arrow] (spinSQS) -- (randomspin);
\draw [arrow] (randomspin) -- (runvasp);
\draw [arrow] (runvasp) -- (end);
\end{tikzpicture}
}
\caption{A DFT-ATAT compatible workflow for computing paramagnetic total energies}
\label{fig:flowchart}
\end{figure}
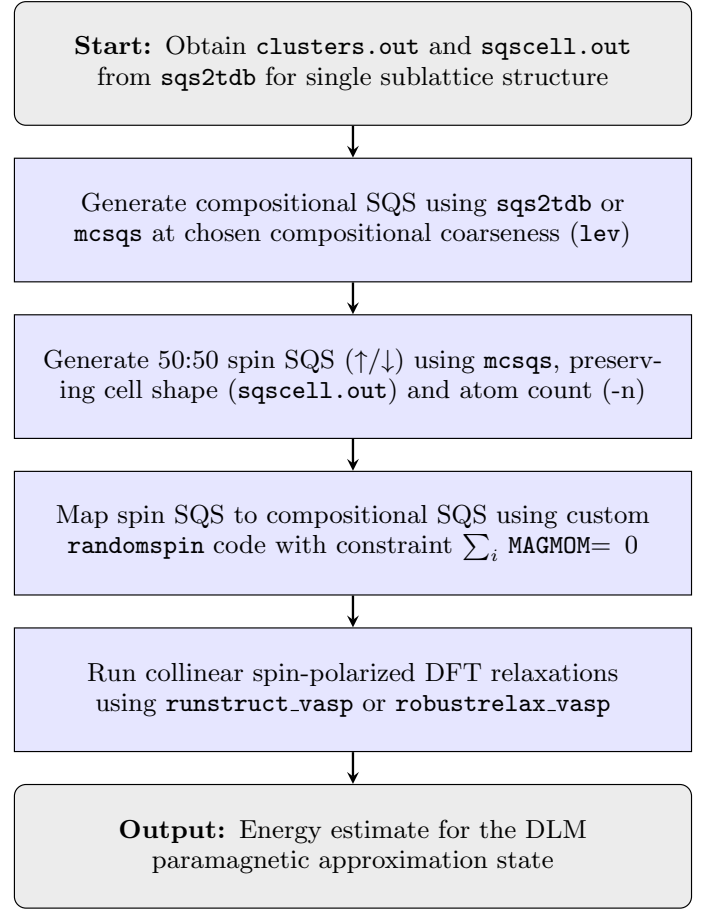

\begin{figure}[pos=!htbp]
\centering
\includegraphics[width=1\linewidth]{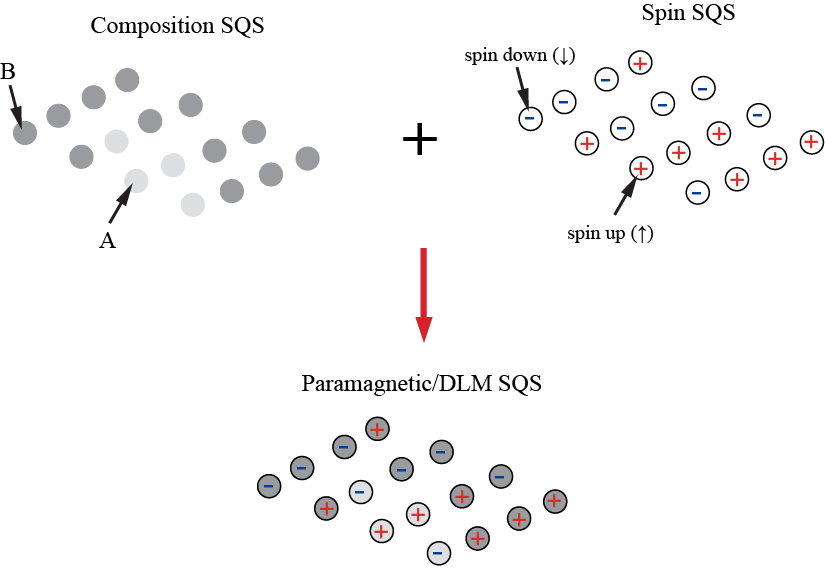}
\caption{Schematic of the adapted DLM-SQS Method for an FCC A = 37.5\%, B = 62.5\% binary alloy}
\label{fig:paramag}
\end{figure}

In this work, we take steps towards simplifying and automating this method via ATAT, as outlined in Figure~\ref{fig:flowchart}. A compositionally disordered SQS is first generated~\cite{vandeWalle2009}, followed by a separate 50:50 spin SQS constructed with \texttt{mcsqs} using the same supercell shape, atomic positions and atom count. In principle, chemical and spin disorder could be optimized simultaneously by generating a four-component SQS (e.g. Ni$^{+}$, Ni$^{-}$, Co$^{+}$, Co$^{-}$); however, matching the correlation functions of a four-component random alloy requires substantially larger supercells. We find it sufficiently accurate to instead overlay two independently optimized two-component SQSs, with the spin pattern applied under a lattice translation relative to the chemical configuration; the shift decorrelates the chemical and spin decorations to a sufficient extent while retaining the smaller cells of the binary construction. The overlay is performed by a custom \texttt{randomspin} code, which assigns the up/down spin pattern to atomic sites and writes the result to the \texttt{str.out} file that \texttt{runstruct\_vasp} reads to generate VASP input files for collinear spin-polarized relaxation calculations (specifically, \texttt{POSCAR} and the \texttt{MAGMOM} line of the \texttt{INCAR}). The code also checks that each chemical species individually receives an equal number of spin-up and spin-down assignments, yielding net zero magnetization both per species and globally, the latter further enforced during electronic optimization by the \texttt{NUPDOWN = 0} tag in VASP. The composition SQS, spin-disorder SQS and combined paramagnetic/DLM SQS are shown schematically in Figure~\ref{fig:paramag} for an A-B binary alloy of A = 37.5~at.\% for reference. For this study of the Co-Ni binary, we construct 16-atom FCC and HCP spin-composition SQSs with initial magnetic moments of $\pm 2\,\mu_B$ per Co and Ni atom. This approach reduces statistical noise and ensures thermodynamic consistency with the paramagnetic state. All spin-polarized DFT relaxations are performed with VASP using parameters detailed in Section~\ref{sec:firstprinciple}.\smallskip

\subsubsection{Estimating Magnetic Transition Temperatures: a DFT CALPHAD-compatible framework}
\label{sec:fcc_coni}

This section outlines how the magnetic transition temperature, i.e., Curie temperature $(T_C)$ for a ferromagnetic system, and can be approximated as an explicit function of the free energy difference between paramagnetic and ferromagnetic states and the effective magnetic moment of the paramagnetic state, $\beta_{\rm DLM}$.\smallskip

Under the constraints of Heisenberg's Model, we can determine that magnetic transition temperature scales directly with $J$, the exchange coupling strength of magnetic interactions. Specifically, considering the Heisenberg Hamiltonian on a lattice \cite{stanley1971introduction,yeomans1992statistical}:
\begin{equation}
\mathcal{H} = -J \sum_{\langle i,j \rangle} \vec{S}_i \cdot \vec{S}_j
\label{eq:hamiltonian}
\end{equation}
where $J$ is the exchange coupling strength and $\langle i,j \rangle$ denotes nearest-neighbor pairs, we can use the canonical partition function to establish the linear scaling of $T_C \propto J$ (See Appendix \ref{sec:IsingHamiltonian} for detailed proof).\smallskip

Within the same constraints of Heisenberg's Model, the following explicit expression can be derived for exchange coupling strength J in terms of the energy difference between the paramagnetic and ferromagnetic states:
\begin{eqnarray}
    Jz = \frac{2(E_{DLM} - E_{FM})}{\mu^2}
    \implies \left(E_{DLM} - E_{FM}\right) \propto T_C
\end{eqnarray}
The DLM paramagnetic state is used here for both the energy reference and the effective magnetic moments, and these two quantities are inherently coupled: the local moments that emerge from the DLM calculation reflect the same disordered magnetic environment that determines $E_{\mathrm{DLM}}$. Using $E_{\mathrm{AFM}}$ in place of $E_{\mathrm{DLM}}$ would decouple the energy reference from the moment calculation, introducing a thermodynamic inconsistency. The intuitive approximation of $\frac{1}{2}(E_{\mathrm{AFM}} - E_{\mathrm{FM}}) \approx E_{\mathrm{DLM}} - E_{\mathrm{FM}}$ assumes a symmetric energy landscape about the DLM state, but the scaling between these energy differences is structure-dependent within the Curie--Weiss framework. Additionally, $E_{\mathrm{AFM}}$ itself is not uniquely defined as multiple antiferromagnetic orderings exist for a given crystal structure (e.g., AFM-I vs.\ AFM-II in BCC), each yielding different energies, whereas the SQS-based DLM construction provides a single, well-defined approximation to the paramagnetic state in a ferromagnetic alloy.\smallskip

\paragraph{Energy Scaling}
\label{sec:energy-scaling}

\begin{figure*}[pos=!tp]
    \centering
    \includegraphics[width=0.8\linewidth]{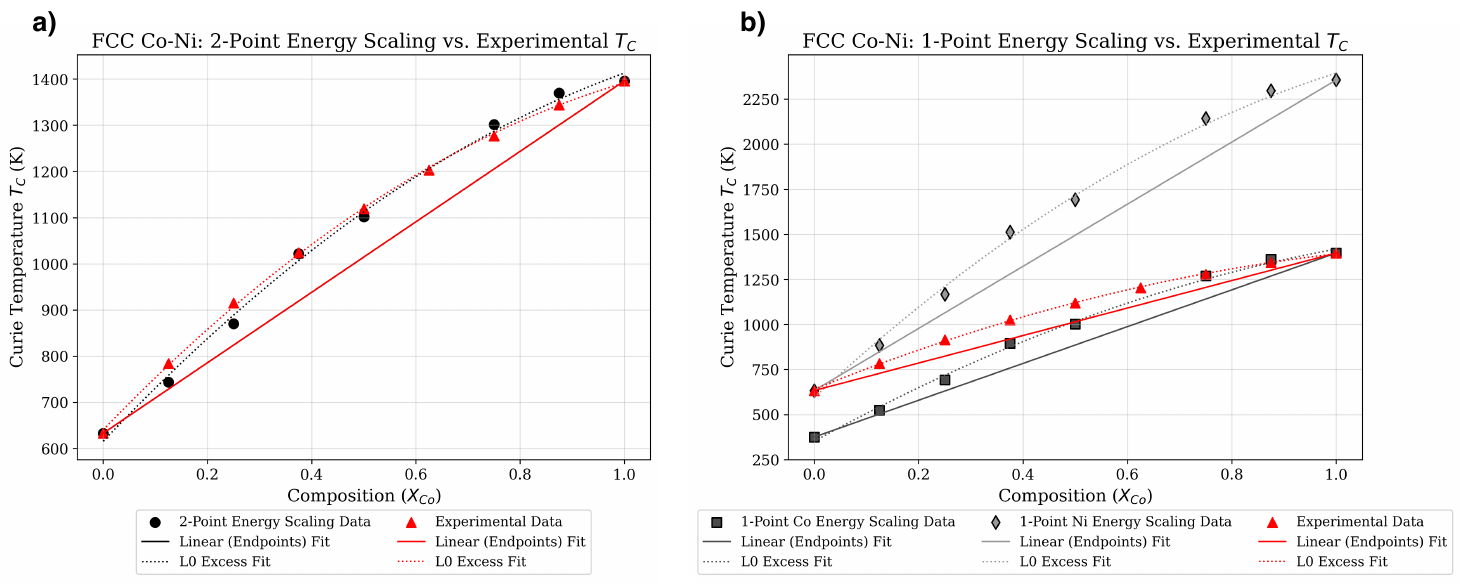}
    \caption{\textbf{a}) $E_{\rm DLM}-E_{\rm FM}$ is directly scaled using the known $T_C$ of Co and Ni FCC endmembers to obtain $T_C$ of the $\rm Co_{x} \rm Ni_{1-x}$ FCC binary alloy, plotted here as a function of $x$ ($X_{\rm Co}$). \textbf{b}) $E_{\rm DLM}-E_{\rm FM}$ is directly scaled using one each of the known $T_C$ of Co and Ni FCC endmembers to obtain $T_C$ of the $\rm Co_{x} \rm Ni_{1-x}$ FCC binary alloy, plotted here as a function of $x$ ($X_{\rm Co}$). Results from known experimental $T_C$ values \cite{FernandezGuillermet} compared for both a) and b).}
    \label{fig:all E scaling}
\end{figure*}

With this knowledge, we can directly scale the energy difference between the paramagnetic and ferromagnetic states with endpoints fixed to the known Curie temperatures of endmembers obtained from SGTE. That is given that $\left(E_{DLM} - E_{FM}\right) = E_{\rm DLM-FM} \propto T_C$, we propose the following 2-point scaling for an $A_xB_{1-x}$ alloy:
\begin{equation}
\begin{split}
    T_C^{\rm 2pt}(x)=T_C^{\rm SGTE}(1)\frac{E_{\rm DLM-FM}(x)-E_{\rm DLM-FM}(0)}{E_{\rm DLM-FM}(1)-E_{\rm DLM-FM}(0)} + \\
    T_C^{\rm SGTE}(0)\frac{E_{\rm DLM-FM}(1)-E_{\rm DLM-FM}(x)}{E_{\rm DLM-FM}(1)-E_{\rm DLM-FM}(0)}
\end{split}
\end{equation}
To validate this formula, the Curie temperatures of the FCC ${\rm Co}_x{\rm Ni}_{1-x}$ binary alloy computed via this method are plotted in Figure \ref{fig:all E scaling}a as a function of $X_{\rm Co}$, with experimental Curie temperatures also plotted for comparison. We observe here that the computed $T_C$ agrees very well with experimental $T_C$, such that the average error is $\pm 22K$ for intermediate compositions, with a maximum error of $40-45K$ under experimental values at $12.5\%-25\%$ Co.\smallskip

However, given that our reason for needing to compute Curie temperature for CALPHAD is to find magnetic parameters at the unknown endmember of HCP Ni, we need an approach that works even if we have only one known Curie temperature. We thus propose the following two alternatives,\smallskip
\begin{eqnarray}
    T_C^{\rm 1pt}(x)=T_C^{\rm SGTE}(0)\frac{E_{\rm DLM-FM}(x)}{E_{\rm DLM-FM}(0)}\\
    T_C^{\rm 1pt}(x)= T_C^{\rm SGTE}(1)\frac{E_{\rm DLM-FM}(x)}{E_{\rm DLM-FM}(1)}
\end{eqnarray}

In Figure \ref{fig:all E scaling}b, we observe one-point normalization with Co and Ni separately for the FCC ${\rm Co}_x{\rm Ni}_{1-x}$ binary alloy, i.e., from the SGTE values of $T_C^{\rm SGTE}(1) = 1396 K$ and $T_C^{\rm SGTE}(0) = 633K$, each independently considered. For the Co-normalized Curie temperatures, while the Curie temperatures for the FCC Co-Ni binary between $75\%-100\%$ Co agree reasonably well ($\pm$ 12.4K) with experimental data, the Curie temperatures for 50\% and below are 110-260K lower (average 197K) than corresponding experimental values. The average Curie temperature error is 144.54K for these results. Overall, the error from experimental measurements increases proportional to the increasing composition of Ni in the binary. For the Ni normalized Curie temperatures, these errors are significantly larger, with values being overestimated by an average of 599K across all compositions, with a maximum error at pure Co of 960K. We see here as well that the error from experimental measurements increases proportional to the increasing composition of Co in the binary, albeit at a much larger magnitude (higher proportionality constant/slope of increase in error). In summary, it is established that compared to the 2 point normalized method, the 1-point normalization is not a good approximation and is not reliable to get values of Curie temperature for metastable phases or where SGTE data is lacking. Furthermore, there is no simple one-point scaling scheme that will predict physical values of Curie temperature. Thus, we ask:
\begin{enumerate}
    \item Can Curie temperature be directly calculated using first principles data, that is, with no normalization?
    \item Are these Curie temperatures better suited for a one-point normalization scheme?
\end{enumerate}

To address our first question, we must turn back to the earlier discussion of thermodynamics used to produce the explicit expression for exchange coupling strength $J$ in terms of $E_{\rm DLM-FM}$. Two approaches for obtaining $T_C$ from first-principles energies are considered here, both within the mean-field approximation but differing in their treatment of the magnetic transition.

\paragraph{`Spin-Counting' Method}
\label{sec:spin-counting}

In the first approach, which we term the `Spin-Counting' (SC) approach, we assume a first-order magnetic transition, i.e., the ordered and disordered free energies are set equal at $T_C$, and the FM state is taken to have negligible magnetic entropy relative to the fully disordered paramagnetic state (see Appendix~\ref{sec:Smag_derivation} for detailed derivation). Under these assumptions, along with the approximation that non-magnetic contributions to the energy difference are negligible and that the DFT energies capture temperature-independent exchange coupling, the transition temperature reduces to
\begin{equation}
    T_C \approx \frac{E_{\mathrm{DLM}}-E_{\mathrm{FM}}}{S_{\mathrm{mag}}}
\end{equation}
where $S_{\mathrm{mag}}$ is the maximum magnetic entropy of the fully disordered paramagnetic state. The entropy itself is obtained by counting the number of accessible orientational micro-states. For quantum spin $S$, the number of spin states is $\Omega = 2S+1$, giving $S_{\mathrm{mag}}^{\mathrm{QM}} = k_B \ln{(2S+1)}$~\cite{Brannvall2024} where $S = \mu = \frac{\beta_{\rm DLM}}{2}$. Thus, we obtain from first principles (fp) calculated values the following generalized expression for magnetic entropy and, as a result, the Curie temperature for an $A_xB_{1-x}$ alloy :
\begin{eqnarray}
    S_{\mathrm{mag}} = k_B \ln{(\beta_{\mathrm{DLM}} + 1)}\\
    \implies T_C^{\mathrm{fp, SC}}(x) = \frac{E_{\mathrm{DLM}}(x)-E_{\mathrm{FM}}(x)}{k_B\ln{(\beta_{\mathrm{DLM}}(x)+1)}}
    \label{eq:MFA}
\end{eqnarray}
This expression assumes only that the paramagnetic state is fully disordered, with each spin orientation equally probable and independent of all others. The entropy is a purely combinatorial quantity, and the first-order transition assumption that magnetic entropy changes discontinuously at $T_C$ is technically incorrect for the second-order ferromagnetic/paramagnetic transition observed in most metallic systems.\smallskip

\paragraph{`Curie-Weiss' Method}
\label{sec:curie-weiss}

Alternatively, the `Curie-Weiss' (CW) approach derives $T_C$ directly from the self-consistent mean-field equations without assuming a first-order transition. Each spin interacts with an effective molecular field produced by its neighbors, and the transition temperature emerges as the point at which the paramagnetic solution becomes unstable to spontaneous magnetization. The generalized maximum magnetic entropy and Curie temperature for an $A_xB_{1-x}$ alloy follow from the critical condition of the self-consistent partition function (see Appendix~\ref{sec:beta_eff_TC} for detailed derivation):
\begin{eqnarray}
    S_{\mathrm{mag}} = k_B\frac{3\beta_{\mathrm{DLM}}}{\beta_{\mathrm{DLM}}+2}\\
    \implies T_C^{\mathrm{fp, CW}}(x) = (E_{\mathrm{DLM}}(x)-E_{\mathrm{FM}}(x))\frac{\beta_{\mathrm{DLM}}(x)+2}{3k_B\beta_{\mathrm{DLM}}(x)}
    \label{eq:CW}
\end{eqnarray}
The Curie-Weiss expression is physically preferable for two reasons. First, it correctly treats the magnetic transition as second-order, consistent with the continuous nature of the ferromagnetic/paramagnetic transition. Second, the entropy is derived from the same Hamiltonian that governs the magnetic ordering, ensuring internal thermodynamic consistency between the energy and entropy used to estimate $T_C$. In the Spin-Counting approach, by contrast, the entropy depends only on the number of accessible states and is therefore insensitive to the nature or strength of the magnetic interactions.\smallskip

\begin{figure}[pos=!htbp]
    \centering
    \includegraphics[width=1\linewidth]{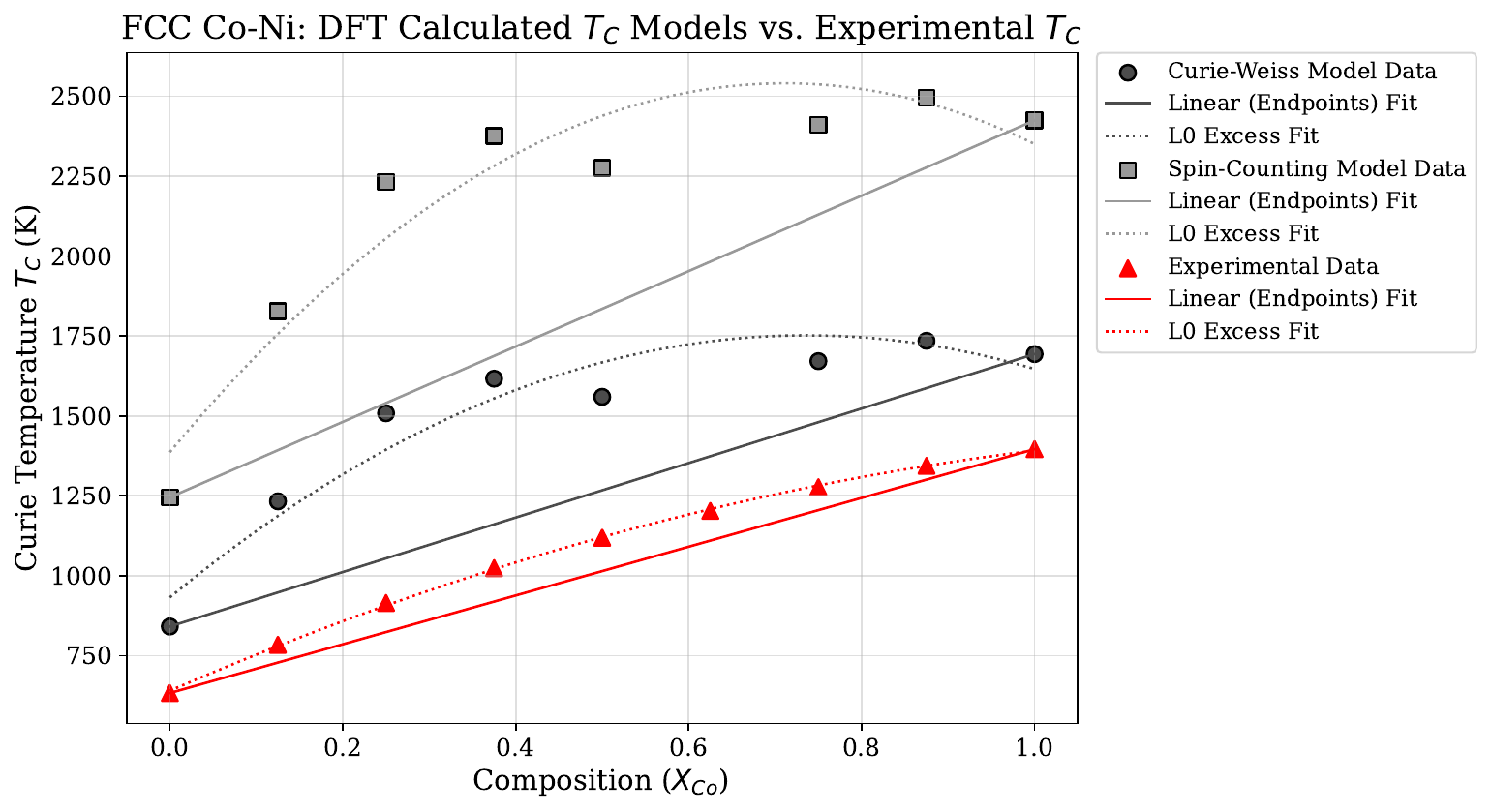}
    \caption{$T_C$ directly calculated using first-principles DFT data (no scaling) from equations derived from the Curie-Weiss model and the Spin-Counting model have been compared for the $\rm Co_{x} \rm Ni_{1-x}$ FCC binary alloy, plotted here as a function of $x$ ($X_{\rm Co}$). Known experimental $T_C$ values \cite{FernandezGuillermet} compared.}
    \label{fig:Calc T_C}
\end{figure}

Our results further support this physical assessment. Comparing the Spin-Counting model and the Curie-Weiss model for the FCC ${\rm Co}_x{\rm Ni}_{1-x}$ binary in Figure \ref{fig:Calc T_C}, we see that while Curie temperature deviates significantly from experimental values for both models, the Curie-Weiss model agrees better with experimental results compared with the Spin-Counting model. The average error for the Curie Weiss Model is 420.72K above experimental values, while the average error for the Spin-Counting model is approximately 1099.35K above experimental values.\smallskip

\paragraph{$\beta_{\rm DLM}$ Correction}
\label{sec:beta-corr}

With knowledge of the Curie-Weiss model performing better and being better motivated theoretically than that the Spin-Counting model, are there any modifications that can be made that would allow better agreement with experimental results and the energy scaling 2-point normalization method? Turning to a thermodynamic/statistical mechanics perspective, is there a method to implicitly correct the magnetic entropy term that accounts for all assumptions we make regarding the nature of local magnetic moments extracted from the paramagnetic SQS? Particularly, for the FCC ${\rm Co}_x{\rm Ni}_{1-x}$ binary, the effective magnetic moments are less than 1 for 0$\%$-62.5$\%$ Co, and in this same region, there are slope irregularities (parabolic behavior) in the calculated Tc trends that are not commensurate with experimental trends. \smallskip

We know that for localized magnetic moments with spin quantum number S, the magnetic moment is $\beta_{\rm DLM} = 2S$. Here, spin S can be 0, 1/2, 1, 3/2, etc. While S = 0 corresponds to a nonmagnetic state with a single spin state and therefore no magnetic entropy, the lowest nonzero spin, $S = \frac{1}{2}$, gives $\beta_{\rm DLM} = 2S = 1$. If magnetic entropy is to be included at all, the counting assumes at least 2 spin states per atom ($S \ge \frac{1}{2}, \beta_{\rm DLM} \ge 1$), so a calculated moment with $\beta_{\rm DLM} < 1$ has no counterpart in this localized-spin counting. More rigorously, we can justify our correction of $\beta_{\rm DLM} < 1$ as a universal scaling across all betas as correcting for Short Range Order. When short-range order (SRO) is neglected, we treat the spin orientations as statistically independent and count only the possible spin states at each site, but SRO effects seemingly dampen local magnetic moments of the SQS post VASP relaxation of collinear calculations. Therefore, the following correction of $\beta_{\rm DLM}$ is proposed:
\begin{equation}
    \beta^{\rm corr}_{\rm DLM} = \beta_{\rm DLM} + A\exp(-C\beta_{\rm DLM})
    \label{eq:beta_corr}
\end{equation}

The constants in Eq.~(\ref{eq:beta_corr}) are not independent, and both follow
from the $S = \frac{1}{2}$ argument above. The amplitude is fixed by the
$S = \frac{1}{2}$ floor, $A = 1$, so that a fully collapsed moment is restored
to the two-state value rather than to an arbitrary one. The decay constant
follows from requiring that the correction never return a moment below that
same floor. Since
\begin{equation}
    \frac{{\rm d}\beta^{\rm corr}_{\rm DLM}}{{\rm d}\beta_{\rm DLM}}
    = 1 - C\exp(-C\beta_{\rm DLM}),
    \label{eq:beta_corr_deriv}
\end{equation}
the corrected moment is monotone, and hence bounded below by its value at the
origin, if and only if $C \le 1$. For $C > 1$ the map turns over at small
$\beta_{\rm DLM}$ and falls below unity, violating the floor it was introduced
to enforce; at $C = 2$, for instance, the corrected moment falls below unity for $\beta_{\rm DLM} < 0.80$, which is half of the moments calculated across the two phases. The remaining requirement is localization: the correction decays over a range $1/C$ and should be confined to the weak-moment region where the collinear DLM moment is unreliable, which favors the largest admissible decay constant. Values $C<1$ are admissible but spread the correction into the large-moment regime where none is warranted. The two conditions together select $C = 1$, equivalently the unique value for which Eq.~(\ref{eq:beta_corr_deriv}) vanishes at the origin, giving
\begin{equation}
    \beta^{\rm corr}_{\rm DLM} = \beta_{\rm DLM} + \exp(-\beta_{\rm DLM}).
    \label{eq:beta_corr_final}
\end{equation}
Equation~(\ref{eq:beta_corr_final}) is thus a one-parameter correction constrained by the spin floor, not a two-parameter empirical fit. Its sensitivity is reported in the Supplemental Information. Because $\beta$ enters the magnetic Gibbs energy only through the prefactor $\ln(\beta^{\rm corr}_{\rm DLM} + 1)$, the fractional change in the magnetic contribution is independent of both temperature and the shape function, and over $C = 0.5$-$2$ it does not exceed $16\%$ for the calculated moments, which span $0.22 \le \beta_{\rm DLM} \le 1.14$. This bounds the effect on the fitted description from above, since the excess parameters are refitted and reabsorb part of any such change (see Fig.~S1 for sensitivity analysis).\smallskip

\begin{figure}[pos=!htbp]
    \centering
    \includegraphics[width=1\linewidth]{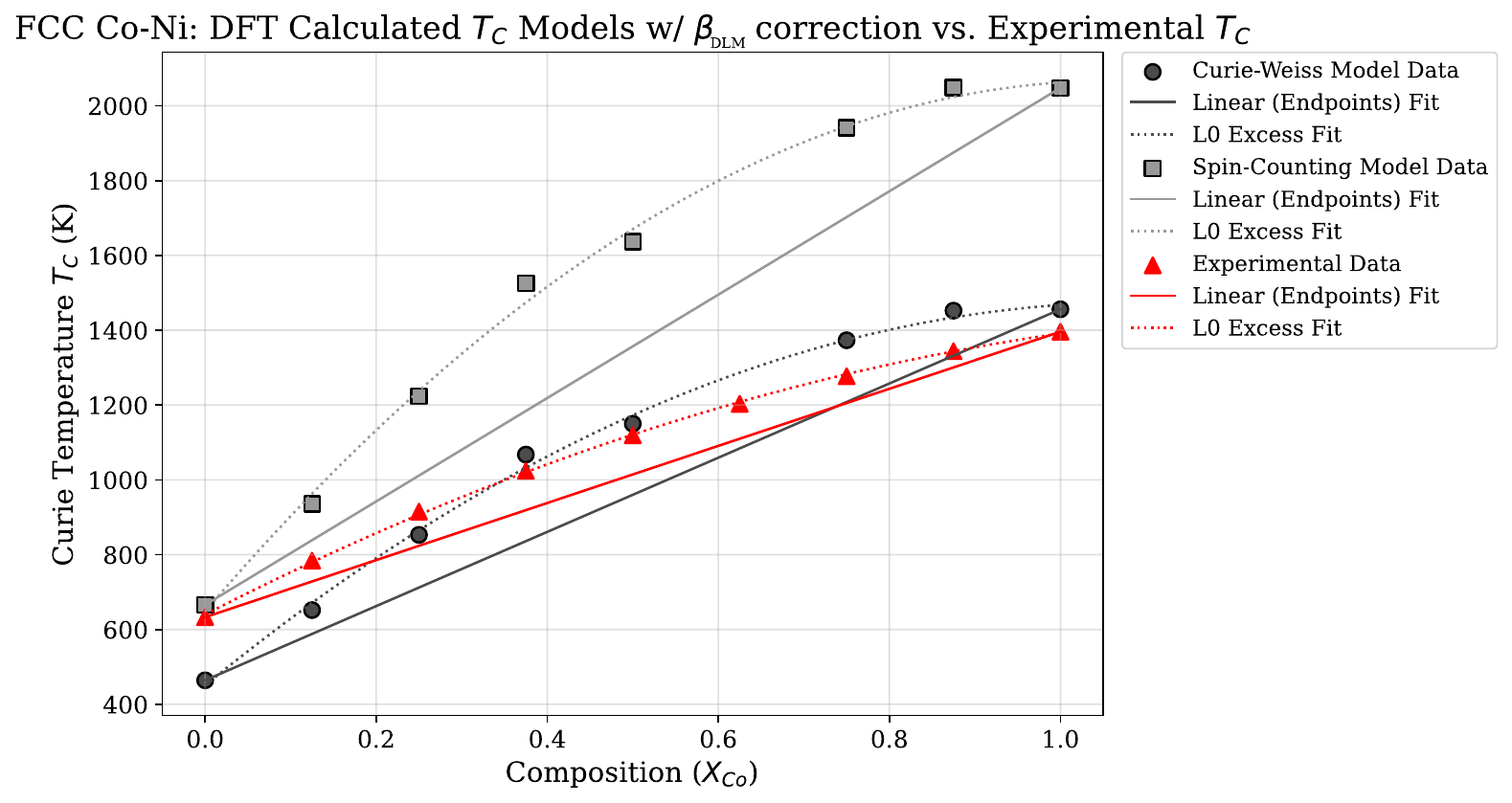}
    \caption{$T_C$ directly calculated using first-principles DFT data (no scaling) from equations derived from the Curie-Weiss model and the Spin-Counting model, along with corrected values from using $\beta^{\rm corr}_{\rm DLM}$ according to Equation \ref{eq:beta_corr}, have been compared for the $\rm Co_{x} \rm Ni_{1-x}$ FCC binary alloy, plotted here as a function of $x$ ($X_{\rm Co}$). Known experimental $T_C$ values \cite{FernandezGuillermet} compared.}
    \label{fig:Calc T_C beta_eff}
\end{figure}

In Figure \ref{fig:Calc T_C beta_eff}, the $\beta_{\rm DLM}$ correction has been applied to both the Curie Weiss model and the Spin-Counting model. Most notable here is the vast improvement in agreement of the calculated $T_C$ with experimental data compared to previous attempt, with an average error of 87.76 K for the Curie Weiss model and 441.73 K for the Spin-Counting model, compared to the previous values of 420.72 K and 1099.35 K respectively.\smallskip

\paragraph{Scaling vs. Direct Calculation Performance}
\label{sec:svaling-v-direct}

But can we leverage the known Curie temperatures of one member to provide reasonable extrapolations for the other and good agreement overall with experimental measurements? For this, we propose the following formula similar to the earlier 1-point normalization for energy scaling, specifically the normalized Curie temperature at any composition $x$ can be approximated as:

\begin{eqnarray}
T_C(x) = T_C^{\text{SGTE}}(0) \cdot \frac{T_C^{\text{fp}}(x)}{T_C^{\text{fp}}(0)}\\
T_C(x) = T_C^{\text{SGTE}}(1) \cdot \frac{T_C^{\text{fp}}(x)}{T_C^{\text{fp}}(1)}
\label{eq:Tc_one_endmember}
\end{eqnarray}

\begin{figure}[pos=!htbp]
    \centering
    \includegraphics[width=1\linewidth]{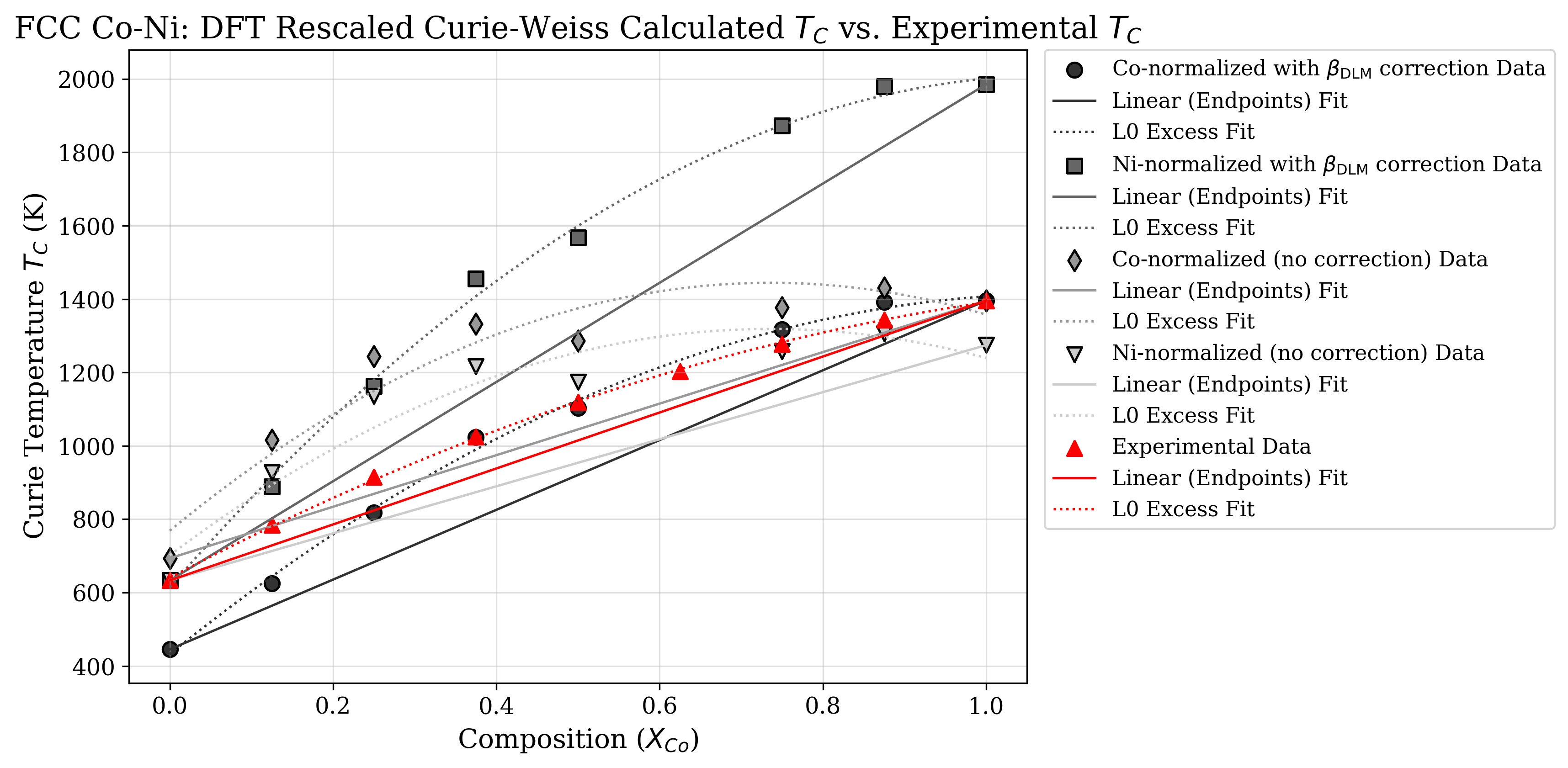}
    \caption{$T_C$ directly calculated from first principles DFT data (for both corrected and uncorrected $\beta_{\rm DLM}$ using the equation derived from the Curie-Weiss model have been scaled with one each of the known Co and Ni FCC endmember $T_C$, plotted here as a function of $x$ ($X_{\rm Co}$). Known experimental $T_C$ values \cite{FernandezGuillermet} compared.}
    \label{fig:Rescale T_C}
\end{figure}

Figure~\ref{fig:Rescale T_C} reveals key limitations of the one-point normalization scheme. When using the corrected $\beta^{\rm corr}_{\rm DLM}$, the predicted $T_C$ is highly sensitive to the choice of normalizing endmember. Specifically, Co-normalized values agree reasonably well with experiment (average error of
78.41~K), while Ni-normalized values deviate significantly (average error of 436.43~K). When using the uncorrected $\beta_{\rm DLM}$, the sensitivity to the normalizing endmember is substantially reduced. However, the irregular trends in $T_C$ vs.\ composition discussed in the context of the $\beta_{\rm DLM}$ to $\beta^{\rm corr}_{\rm DLM}$ correction persist. Thus, both normalizations perform reasonably only in Co-rich compositions, with accuracy degrading substantially in the Ni-rich region.\smallskip

Overall, the directly calculated (unnormalized) $T_C$ using $\beta^{\rm corr}_{\rm DLM}$ yields the lowest average error against experiment among all methods except the energy-scaling two-point normalization. The two-point scheme can only be used when both endmember $T_C$ values are known, but the proposed Curie-Weiss approach with effective moment corrections provides the next most accurate results and remains applicable when one or both endmember values are experimentally unavailable. However, while the IHJ model produces the full magnetic heat capacity from $T_C$ and $\beta$ alone, the $\beta$ values in established CALPHAD assessments~\cite{FernandezGuillermet} are not bare magnetic moments but effective parameters fitted to reproduce experimental thermodynamic data. These values implicitly compensate for limitations to the mean-field IHJ framework, including short-range order above $T_C$ and the simplified shape of the lambda anomaly. DLM-derived magnetic moments, though physically well-defined, do not carry these empirical corrections and therefore yield different magnetic entropy contributions when used directly in the IHJ model. Because the one-point normalization is unreliable in both cases, and because raw DLM-derived magnetic moments alone contribute systematic error in heat capacity and entropy when compared to Guillermet's experimentally fitted parameters, we
turn in Section~\ref{sec:results} to a prior-informed
optimization that anchors the Redlich-Kister coefficients to
established CALPHAD values while allowing the DFT data to
determine the composition dependence.\smallskip

\begin{figure}[pos=!htbp]
    \centering
    \includegraphics[width=1\linewidth]{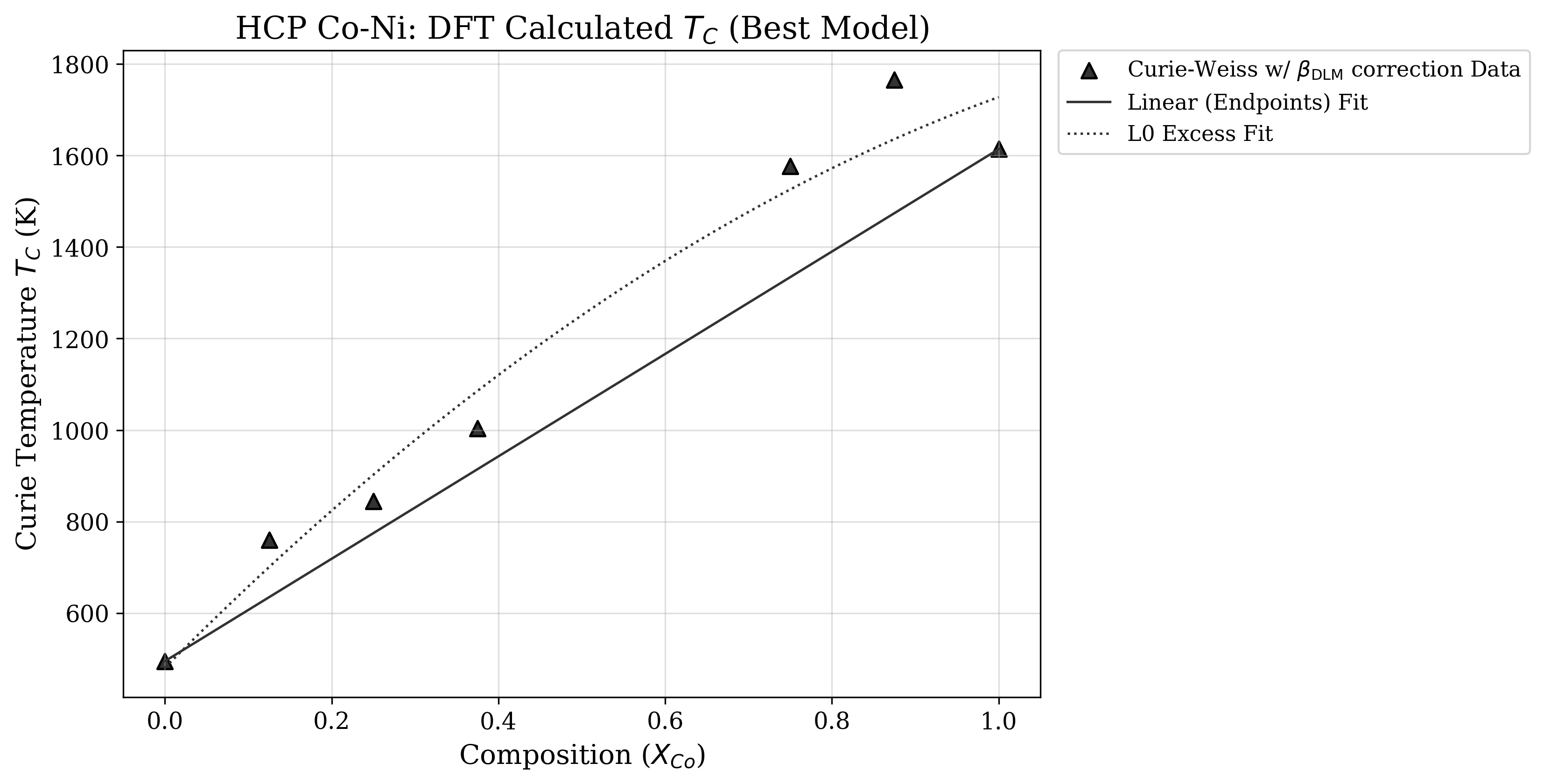}
    \caption{$T_C$ directly calculated using first-principles DFT data (no scaling) from the equation derived from the Curie-Weiss model applying correction for $\beta_{\rm DLM}$ (Equation \ref{eq:beta_corr}) for the $\rm Co_{x} \rm Ni_{1-x}$ HCP binary alloy, plotted here as a function of $x$ ($X_{\rm Co}$).}
    \label{fig:hcp_calc}
\end{figure}

To obtain the Curie temperature for HCP Ni, the analysis conducted for FCC Co--Ni is repeated for HCP Co--Ni, with the most significant results according to our prior findings being shown in Figure~\ref{fig:hcp_calc}.\smallskip

\subsubsection{Estimating the $\beta$ parameters from DFT Magnetic Moment ($\mu_B$) Data}\label{sec:magmom}

\begin{figure}[pos=!htbp]
    \centering
    \includegraphics[width=1\linewidth]{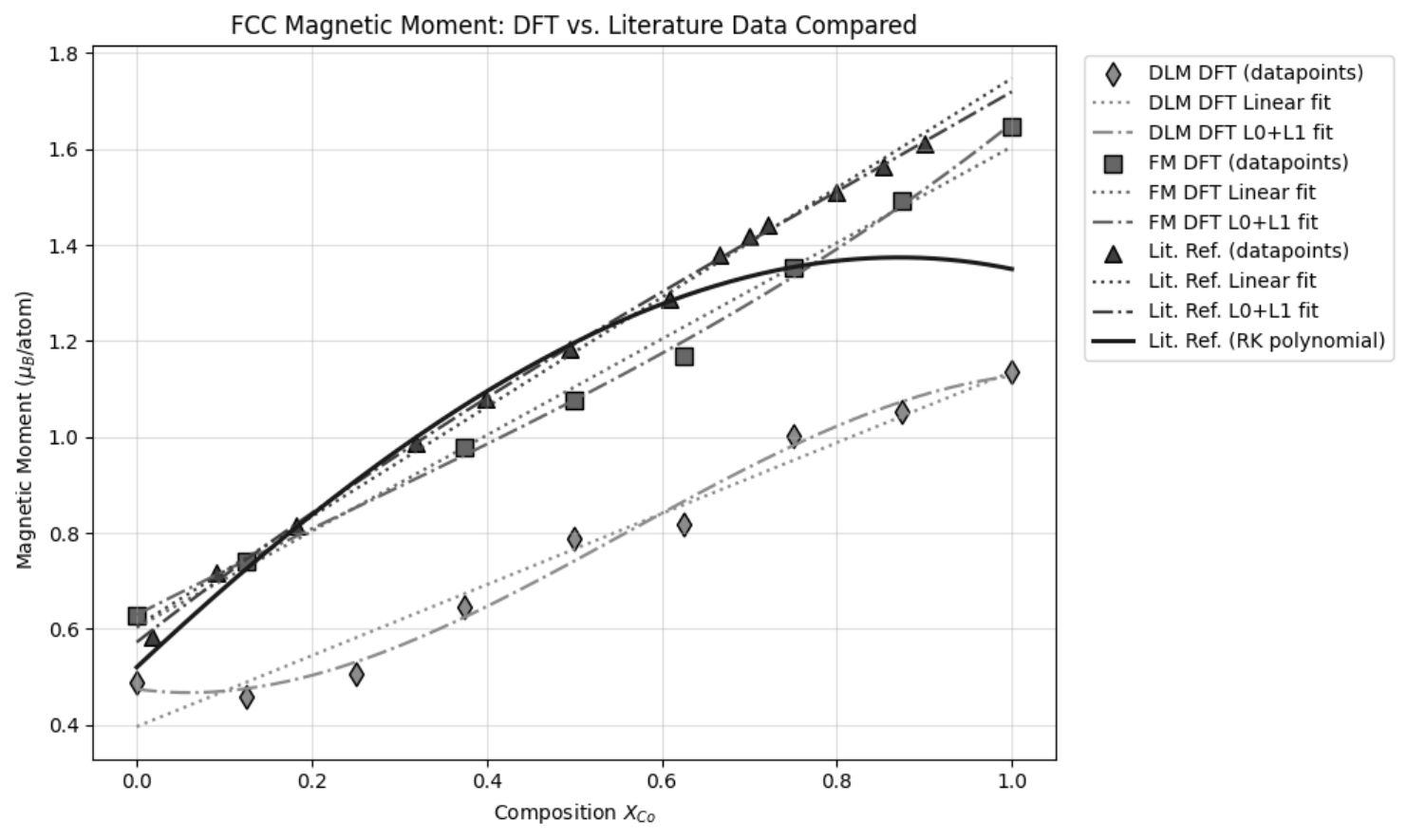}
    \caption{Effective magnetic moments calculated from DLM and FM SQS relaxations across the FCC Co-Ni binary}
    \label{fig:BMAG-FCC}
\end{figure}

In our earlier discussion of the Inden-Hillert-Jarl model in Section \ref{sec:magneticmodels}, we discussed that the $\beta$ parameter represents the geometric average magnetic moment more consistent with the definition of entropy. A ferromagnetic (FM) DFT calculation determines the ground-state electronic structure at 0K with all spins aligned. The resulting magnetic moment should be the saturation magnetization, i.e., the maximum moment achievable when thermal fluctuations are absent and all exchange interactions are fully satisfied. This assertion aligns with results outlined in Figure \ref{fig:BMAG-FCC}, where we observe that experimental saturation magnetization agrees well with magnetic moment obtained for FM calculations across the FCC Co-Ni binary composition space, differing by a maximum of ~0.1 $\mu_B$.\smallskip

The DLM calculations representing the disordered state, on the other hand, directly model the physical regime that $\beta$ is meant to describe: the paramagnetic phase above $T_C$. The local moment obtained from a DLM calculation is the moment that persists when long-range order is destroyed by thermal fluctuations. It is this moment that determines the number of accessible spin microstates. For the Co-Ni system particularly, the distinction between FM and DLM moments matters as Co and Ni are both itinerant ferromagnets in which the magnetic moment arises from the exchange splitting of d-band electrons rather than from localized atomic moments. In itinerant systems, the local moment is sustained by the
self-consistent exchange splitting, which depends on the
magnetic environment of neighboring atoms. When long-range
order is destroyed, as is in the DLM state, this splitting collapses and the local moment is significantly reduced, as seen in Figure~\ref{fig:BMAG-FCC} where the DLM
moments fall systematically below the FM moments across the
entire composition range \cite{Pindor2000}. Foundational studies \cite{Pindor2000} have performed self-consistent KKR-CPA calculations for Fe, Co, Ni, and Cr in the DLM state and found larger disagreements between paramagnetic and ferromagnetic local moments of Co and Ni (50\% difference, classified closer to `Stoner' itinerant ferromagnets) than are found for Fe (15\% difference, classified closer to localized `Heisenberg' ferromagnets). This distinction is noted in the most widely accepted CALPHAD assessment of the Co-Ni binary as well \cite{FernandezGuillermet}, which states that the simple equivalence between the $\beta$ parameter and ground state (ferromagnetic) magnetic moment in Bohr magnetons per atom that holds true for Fe does not hold for Co and Ni.\smallskip

\begin{figure}[pos=!htbp]
    \centering
    \includegraphics[width=1\linewidth]{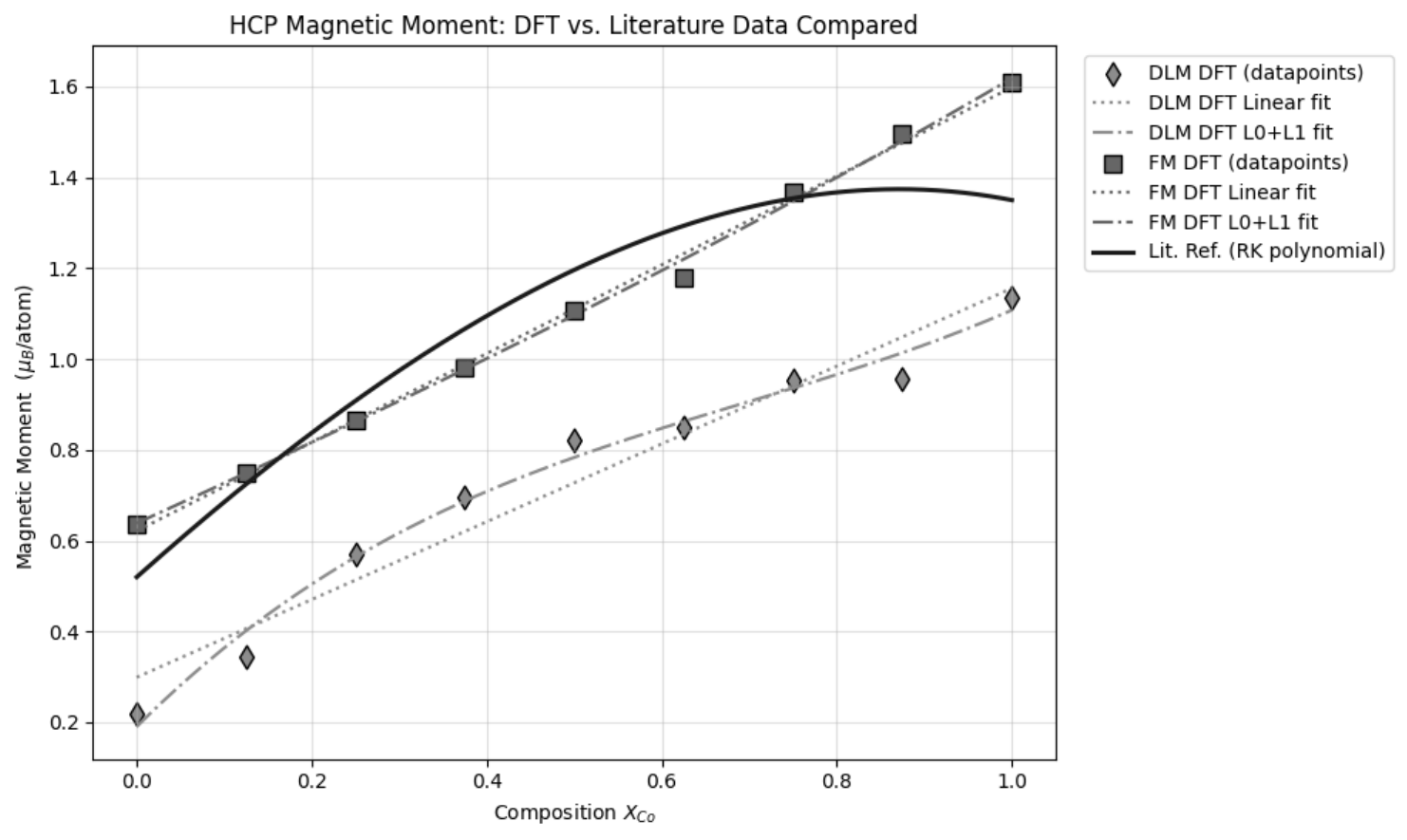}
    \caption{Effective magnetic moments calculated from DLM and FM SQS relaxations across the HCP Co-Ni binary}
    \label{fig:BMAG-HCP}
\end{figure}

The composition dependence of $\beta$ across the Co-Ni binary should therefore be fitted to DLM-derived moments rather than FM moments. Since the FM-to-DLM offset increases
toward the Co-rich side where the absolute moment is largest,
fitting to FM data would not simply shift the RK polynomial
uniformly but would alter its curvature, distorting the
interaction parameters $L_0$ and $L_1$ and overstabilizing
magnetically ordered phases~\cite{Xiong2012}.
Figures~\ref{fig:BMAG-FCC} and~\ref{fig:BMAG-HCP} confirm this
for both FCC and HCP phases, with the FM-DLM discrepancy
reaching approximately $0.5~\mu_\text{B}$ at the Co
endmember.\smallskip \smallskip

The moment obtained for each SQS is the effective moment of Xiong \emph{et al.}~\cite{Xiong2012},
\begin{equation}
    \beta^{*} = \prod_i (\beta_i + 1)^{x_i} - 1,
    \label{eq:beta-effective}
\end{equation}
formed from the local moments resolved by species, and it reproduces the maximum magnetic entropy where a mean over the cell would only approximate it. A non-magnetic species, having $\beta_i = 0$, contributes a factor of unity to Eq.~(\ref{eq:beta-effective}) and enters only by dilution, so alloys mixing magnetic and non-magnetic constituents require no separate
treatment.\smallskip
 
The construction does fix two things about how it is applied. The spin-SQS must randomise orientations only on species carrying a finite local moment, a non-magnetic species being a structural spectator in the magnetic sampling. And the spin floor underlying Eq.~(\ref{eq:beta_corr_final}) is a statement about the number of spin states available to a single local moment, so the correction is applied per species and before combination through Eq.~(\ref{eq:beta-effective}); applied to $\beta^{*}$ directly, or to a non-magnetic species, it would assign $R\ln 2$ of entropy per mole of atoms possessing no local moment.\smallskip

Similarly, we see that the DLM trends also depart from the Guillermet curve for both FCC and HCP, significantly so above $X_{\mathrm{Co}} \approx 0.4$ in Figure~\ref{fig:BMAG-HCP} and above $X_{\mathrm{Co}} \approx 0.5$ in Figure~\ref{fig:BMAG-FCC}. As mentioned earlier in Section~\ref{sec:fcc_coni}, Guillermet's $\beta$ values were not directly measured paramagnetic moments but thermodynamic parameters extracted by fitting the IHJ shape function $f(\tau)$ to experimental heat capacity data measured across the magnetic transition~\cite{FernandezGuillermet}.The resulting $\beta$ therefore encodes information about the $C_p$ anomaly at $T_C$ and the distribution of magnetic entropy across temperature, these quantities being governed primarily by the ordered-state magnetization and the thermodynamics of the continuous transition. This distinction propagates directly into the magnetic entropy and hence into the predicted phase stability at elevated temperatures, and has consequences for how first-principles data should be incorporated into the IHJ framework (Section~\ref{sec:results}).

\section{Results and Discussion}
\label{sec:results}

\subsection{Evaluating Normalization Methods and Determining Best Practices}
\label{sec:normalization}

Now that we have established which sets of first-principles data must be used to obtain the magnetic parameters relating to $T_C$ and $\beta$ for the thermodynamic assessment of the Co-Ni binary, and that observed normalization schemes against one known endmember are unsatisfactory (as seen for the calculation of $T_C$ in the FCC Co-Ni binary in Section \ref{sec:fcc_coni}), we must explore and establish an alternative approach that can be used for both $T_C$ and $\beta$. An identical treatment for these parameters is desirable as they are both used to compute the same magnetic contribution to free energy via the Inden-Hillert-Jarl model. We thus present an approach based on Maximum A Posteriori (MAP) estimation \cite{bishop2006, murphy2012}, in which Gaussian priors derived from experimental and first-principles endmember values, and interaction parameters from established CALPHAD assessments of identical (or similar) systems, regularize the parameter extraction. This approach is analogous to recommendations outlined by textbook CALPHAD optimization methodologies \cite{lukas2007computational}, where successful parameter optimization requires both reliable data and physically informed starting points and bounds. \smallskip

The need for priors is not merely numerical. As discussed in the comparison of DLM-derived and Guillermet $\beta$ parameter values (Section~\ref{sec:magmom}), the IHJ model parameters were originally calibrated against experimental heat capacity data across the magnetic transition \cite{Inden1976, Hillert1978}. The $\beta$ and $T_C$ values in the existing experimental assessment~\cite{FernandezGuillermet} encode how magnetic entropy is distributed as a function of temperature through the shape function $f(\tau)$, which consists of thermodynamic information derived from $C_p$ measurements that our ab-initio results do not provide. Thus, an unconstrained fit to DFT data alone risks yielding IHJ parameters that reproduce the first-principles energetics but are inconsistent with the experimental heat capacity behavior from which the model derives its physical meaning. The priors therefore serve as the channel through which this experimentally measured thermodynamic content enters the
optimization.\smallskip

Numerically, when conventional least-squares fitting is used with one fixed endmember, systematic DFT errors arising from exchange-correlation approximations or basis set convergence \cite{Lejaeghere2014} propagate linearly into every mixing data point and are absorbed into the interaction parameters $L_n^{(P)}$, conflating physical mixing behavior with computational artifacts. The least-squares method works well only when scatter is completely random \cite{lukas2007computational}: for DFT-computed magnetic properties, systematic deviations are inherent to the methodology and cannot simply be excluded. The MAP estimator \cite{bishop2006, murphy2012} addresses this by minimizing:
\begin{equation} 
    \hat{\theta}_{\mathrm{MAP}} = \arg\min{\theta} { \sum_i \frac{\left[y_i - P^{\phi}(x_i;,\theta)\right]^2}{\sigma_y^2} + \sum_j \frac{(\theta_j - \mu_j)^2}{\sigma_j^2}} \label{eq:MAP} 
\end{equation} 
where $y_i$ are the DFT-computed property values at composition $x_i$, $\sigma_y$ is the expected DFT uncertainty, and $\mu_j$, $\sigma_j$ are the prior mean and standard deviation for each Redlich--Kister parameter. The first term enforces fidelity to the DFT data; the second penalizes deviation from prior knowledge, with the ratio $\sigma_y^2/\sigma_j^2$ controlling the effective weight of each prior relative to the data.\smallskip

Conversely, the direct fit uses the same Redlich-Kister form $P^\phi$ and minimizes only the data-misfit term, that is
\begin{equation}
\hat{\theta}_{\mathrm{LS}} = \arg\min_{\theta} \sum_i \left[ y_i - P^\phi(x_i; \theta) \right]^2 
\label{eq:directfit}
\end{equation}
which is Eq.~\eqref{eq:MAP} with the prior penalty removed and is equivalent to taking $\sigma_j \to \infty$. Both direct and MAP fitting methods have the functionality of fitting for endmembers $P_A, P_B$ and the interaction parameters $L_n^{(P)}$ for a single property $P \in \{T_C, \beta\}$ and phase
$\phi$ (see Eq.~\ref{eq:rk}). or fixing $P_A$ and/or $P_B$ while fitting for remaining values. A separate fit is performed for each $(P, \phi)$ pair, with the
truncation order $N$ chosen per dataset.

\begin{figure*}[pos=!tp]
	\centering
	\includegraphics[width=0.725\textwidth]{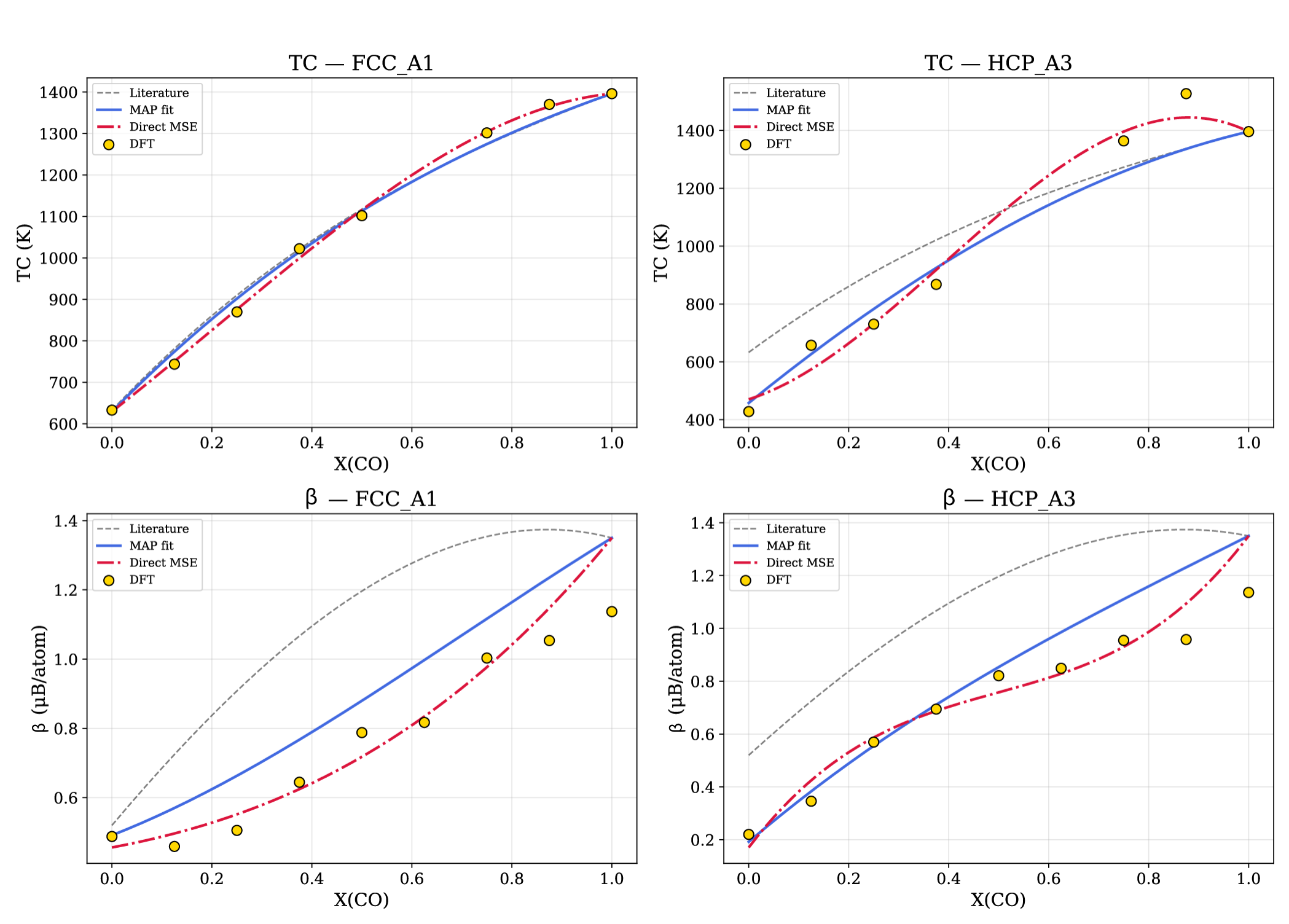}
	\caption{Polynomial fitting of composition dependence of DFT calculated Curie temperature $T_C$ (top row) and magnetic moment $\beta$ (bottom row) for the FCC (left) and HCP (right) phases of the Co-Ni binary. MAP fitting is contrasted with the unconstrained direct least-squares fit (with Co parameters fixed for all fits), and prior literature parameter fitting \cite{FernandezGuillermet} is shown as well. For the FCC phase, both fitting methods produce similar curves, differing primarily in the $L_1$ asymmetry term. For the HCP phase, the direct fit produces nonphysically large $L_1$ values ($L_1^{(T_C)} = 1075$~K, $L_1^{(\beta)} = -1.313$~$\mu_\text{B}$) that create steep inflections not well supported by the DFT data, while the MAP fit yields smooth monotonic curves. Both methods agree that the metastable HCP-Ni endmembers lie well below literature values that were adopted from FCC-Ni.}
	\label{fig:MAP_vs_Direct}
\end{figure*}

The prior means $\mu_j$ and $\sigma_j$ are set according to the experimental support available for each parameter. For the FCC and HCP Co endmembers, the values of $T_C = 1396$~K
and $\beta = 1.35$~$\mu_\text{B}$ are fixed to their SGTE
values~\cite{sgte1991}, equivalent to constraining a ``1-point normalization" reference. Both quantities are well established experimentally and are crucial to maintaining thermodynamic consistency at the Co-rich limit of the Co-Ni binary, particularly the correct HCP-FCC phase transition at
$x_{\mathrm{Co}} = 1$, which has been rigorously validated in
prior CALPHAD assessments~\cite{sgte1991,kaufman2000}. FCC Ni ($P_B$) endmembers receive tight priors ($\sigma = 30$~K for $T_C$, $\sigma = 0.05$~$\mu_\text{B}$ for $\beta$) that permit small adjustments to accommodate the DFT data while remaining anchored to experiment. The metastable HCP Ni endmember, which was assumed equal to FCC Ni in the foundational Guillermet assessment for reasons as mentioned in Section \ref{sec:intro}, receive prior widths set to the magnitude of the discrepancy between the literature assumption and the DFT-computed value (250~K for $T_C$, 0.30~$\mu_\text{B}$ for $\beta$), allowing the DFT data to substantially revise these values while preventing nonphysical results.\smallskip

For interaction parameters, prior standard deviations are set systematically to 50\% of ${L}_0$ for the FCC phase, where the literature assessment is supported by experimental Curie temperature and magnetic moment data across the binary, and to 75\% of ${L}_0$ for the HCP phase, where the interaction parameters were adopted from FCC as outlined in Section \ref{sec:intro}. In both cases, the same absolute $\sigma$ is applied to ${L}_1$ as to ${L}_0$ to avoid artificially constraining the asymmetry term when $|{L}_1| \ll |{L}_0|$. All priors have been listed in Table \ref{tab:magnetic_params}. This hierarchy of prior widths reflects the confidence in data source of each parameter: quantities with direct experimental support are tightly constrained, quantities borrowed from literature of another phase are loosely constrained, and the DFT data determines the balance. The number of SQS compositions computed per phase (7-9) is sufficient to determine the free parameters; the role of the prior is not to compensate for insufficient data but to prevent systematic DFT errors from being absorbed into the interaction parameters.\smallskip

\begin{table*}[pos=!tp]
	\centering
    \renewcommand{\arraystretch}{1.2}
	\setlength{\tabcolsep}{5pt}
	\begin{tabular}{ccl rrrr rrrr}
		\toprule
		& & \textbf{MAP Prior} & \multicolumn{4}{c}{\textbf{MAP Fit}} & \multicolumn{4}{c}{\textbf{Direct Fit}} \\
		\cmidrule(lr){4-7} \cmidrule(lr){8-11}
		\textbf{Property} & \textbf{Phase} & $\mu_j \pm \sigma_j$ & $P_{\rm Ni}$ & $L_0$ & $L_1$ & RMSE & $P_{\rm Ni}$ & $L_0$ & $L_1$ & RMSE \\
		\midrule
		\multicolumn{11}{l}{\textit{Curie temperature $T_C$ (K), \quad $P_{\rm Co} = 1396$~K (fixed, both phases)}} \\
		\midrule
		& & $P_{\rm Ni}$: $633 \pm 30$ & & & & & & & & \\
		$T_C$ & FCC\_A1 & $L_0$: $411 \pm 206$ & $630$ & $402$ & $-56$ & $21.8$ & $629$ & $412$ & $239$ & $10.3$ \\
		& & $L_1$: $-99 \pm 206$ & & & & & & & & \\
		\midrule
		& & $P_{\rm Ni}$: $633 \pm 250$ & & & & & & & & \\
		$T_C$ & HCP\_A3 & $L_0$: $411 \pm 206$ & $458$ & $499$ & $34$ & $89.0$ & $471$ & $697$ & $1075$ & $52.2$ \\
		& & $L_1$: $-99 \pm 206$ & & & & & & & & \\
		\midrule
		\multicolumn{11}{l}{\textit{Magnetic moment $\beta$ ($\mu_\text{B}$), \quad $P_{\rm Co} = 1.35$~$\mu_\text{B}$ (fixed, both phases)}} \\
		\midrule
		& & $P_{\rm Ni}$: $0.52 \pm 0.05$ & & & & & & & & \\
		$\beta$ & FCC\_A1 & $L_0$: $1.046 \pm 0.523$ & $0.491$ & $-0.164$ & $0.131$ & $0.143$ & $0.456$ & $-0.741$ & $-0.117$ & $0.085$ \\
		& & $L_1$: $0.165 \pm 0.523$ & & & & & & & & \\
		\midrule
		& & $P_{\rm Ni}$: $0.52 \pm 0.30$ & & & & & & & & \\
		$\beta$ & HCP\_A3 & $L_0$: $1.046 \pm 0.523$ & $0.192$ & $0.330$ & $-0.125$ & $0.136$ & $0.171$ & $-0.010$ & $-1.313$ & $0.093$ \\
		& & $L_1$: $0.165 \pm 0.523$ & & & & & & & & \\
		\midrule
		\multicolumn{2}{l}{\textbf{DFT Fidelity Score}} & & \multicolumn{4}{c}{$\mathbf{0.804}$} & \multicolumn{4}{c}{$\mathbf{0.920}$} \\
		\bottomrule
	\end{tabular}
    \caption{Redlich-Kister (RK) polynomial magnetic parameters and RMSE of fits for the Co-Ni binary. Here, P is the parameter of either $T_C$ or $\beta$ being subject to RK polynomial fitting. The MAP prior column shows the Gaussian prior mean $\mu_j$ and standard deviation $\sigma_j$ used in Eq.~\ref{eq:MAP}; prior standard deviations for interaction parameters are set to 50\% of $L_0$. Fixed parameters ($\sigma_j \to 0$) are excluded from optimization and held at their SGTE values \cite{sgte1991}, and interaction parameters are taken from literature \cite{FernandezGuillermet}. The DFT noise level was chosen to be $\sigma_y$ is 100~K for $T_C$ and 0.2~$\mu_\text{B}$ for $\beta$. The DFT fidelity score is defined as the mean over all datasets of $\exp(-\mathrm{RMSE}^2 / 2\sigma_y^2)$, yielding a value in $[0,\,1]$ where unity indicates perfect agreement. The MAP fit yields higher RMSE due to regularization of the $L_1$ asymmetry parameters, which prevents overfitting at the cost of a smoother composition dependence.}
    \label{tab:magnetic_params}
\end{table*}

Figure~\ref{fig:MAP_vs_Direct} and Table~\ref{tab:magnetic_params} present the fitted magnetic parameters and their comparison with established literature values. For the Curie temperature, the MAP and direct fits yield nearly identical endmember and $L_0$ values in both phases, confirming that the prior does not distort the fit where the data are informative. The two methods diverge mainly in $L_1$. For FCC $T_C$ the MAP fit gives $L_1 = -56$~K against $239$~K for the direct fit, visible as the steeper upward curvature of the direct fit at intermediate compositions. Both $T_C$ RMSE values lie well within the expected DFT uncertainty of $\sigma_y = 100$~K.\smallskip

The magnetic moment shows the largest departure from the literature curve, concentrated at intermediate compositions. At $x(\mathrm{Co}) = 0.5$ the Guillermet assessment gives approximately $1.20~\mu_B$ against $0.88~\mu_B$ for the MAP fit, a difference of about $0.32~\mu_B$ that approaches $40$ to $50$ percent locally near the peak of the literature curve. This difference is carried almost entirely by $L_0$, which changes sign from $+1.046~\mu_B$ in the prior to $-0.164~\mu_B$ in the MAP fit, reflecting the negative excess curvature of the DLM moments relative to the positive excess of the Guillermet curve.\smallskip

As established in Section~\ref{sec:magmom}, this offset is expected and does not indicate a fitting deficiency. The fitted $\beta$ tracks the DLM moments that the IHJ $\beta$ parameter is meant to represent, whereas the Guillermet $\beta$ is a thermodynamic parameter extracted from heat capacity data across the magnetic transition rather than a directly measured paramagnetic moment, so the two need not coincide~\cite{FernandezGuillermet}. The difference is largest above $x(\mathrm{Co}) \approx 0.5$, where the FM-to-DLM offset is greatest and the local moment is itinerant in character~\cite{Pindor2000}. Fitting $\beta$ to the DLM moments rather than to the Guillermet curve is the choice justified in Section~\ref{sec:magmom}, and the sign and magnitude of $L_0$ follow directly from it.\smallskip

The contrast is more pronounced for the HCP phase, where the unconstrained direct fit produces $L_1$ values that exceed $L_0$ in absolute magnitude for both properties. The resulting curves in Figure~\ref{fig:MAP_vs_Direct} exhibit steep inflections in $T_C$ and $\beta$ between $x = 0.4$ and $0.8$ that are not
well supported by the DFT data at those compositions, and are instead artifacts of an under-constrained asymmetry parameter absorbing scatter in the fit. The MAP fit constrains these asymmetry terms, producing monotonic curves at the cost of modestly higher RMSE (Table~\ref{tab:magnetic_params}). This tradeoff is acceptable: the parameters most affected by the prior ($L_1$ terms) are those with the smallest physical impact on the predicted property curves, since the $L_1$ basis function $x(1-x)(2x-1)$ has a maximum amplitude of only 0.19 compared to 0.25 for $L_0$.\smallskip

A key finding common to both methods is the substantial downward revision of the HCP Ni endmember parameters relative to prior literature assumptions. For $T_C$, both fits place HCP Ni at approximately 460~K, well below the 633~K value adopted from FCC Ni. For $\beta$, both fits give approximately 0.18~$\mu_\text{B}$, less than half the literature value of 0.52~$\mu_\text{B}$. That both methods (one prior-informed, the other unconstrained) independently arrive at these values provides strong evidence that the DFT data, rather than the prior, is driving this revision. Our data points to literature curves in Figure~\ref{fig:MAP_vs_Direct} having systematically overestimated $\beta$ across most of the composition range for both phases, consistent with the use of ferromagnetic saturation moments in the original assessment rather than the DLM paramagnetic moments employed here.\smallskip

\subsection{Independent validation of the HCP parameters}
\label{sec:hcp-validation}

The FCC parameters obtained above can be assessed directly against the experimental Curie temperatures and magnetic moments compiled in Sections~\ref{sec:energy-scaling} and~\ref{sec:beta-corr}. No such assessment is possible for HCP Ni, which is the parameter set this work exists to supply. The only independent first-principles values reported for this endmember are those of Shi et al.~\cite{SHI2025121476}, obtained by a route that shares no step with the present approach: nearest-neighbour Heisenberg exchange constants extracted from ferromagnetic--antiferromagnetic total-energy differences, with $T_{\text{C}}$ taken as the peak of the magnetic specific heat in Monte Carlo simulation.\smallskip

For the Curie temperature the two methods agree closely. We obtain $458$~K against their $535$~K, a difference of $77$~K that is smaller than the $89.0$~K RMSE of the HCP $T_{\text{C}}$ fit itself (Table~\ref{tab:magnetic_params}). More importantly, both revise the endmember sharply downward from the $633$~K adopted from FCC Ni in the Guillermet assessment~\cite{FernandezGuillermet} and carried into widely used databases~\cite{LIU2016125}, by $28\%$ in our study and by $15\%$ in Ref.~\cite{SHI2025121476}. While the agreement between the MAP and direct fits reported above establishes that the revision is driven by the DFT data rather than by the priors, the agreement with an independent method and an independent electronic-structure code establishes that it is not an artifact of the DLM construction either. That the present value lies below theirs is expected as their exchange constants are evaluated with rigid, composition-independent moments and possibly omit the longitudinal moment reduction accompanying the loss of magnetic order, which the DLM reference incorporates and which lowers $T_{\text{C}}$.\smallskip

The magnetic moments are not directly comparable, and the apparent discrepancy between $\beta_{\text{HCP-Ni}} = 0.192~\mu_{\text{B}}$ obtained here and $0.7043~\mu_{\text{B}}$ in Ref.~\cite{SHI2025121476} illustrates the distinction drawn in Section~\ref{sec:beta-corr}. Their value is a ferromagnetic moment, in line with the $0.62~\mu_{\text{B}}$ reported from earlier spin-polarized calculations for HCP Ni, and is inserted into the IHJ model in that form. The present $\beta$ is the DLM paramagnetic moment, which is the quantity the IHJ model's $R\ln(\beta+1)$ term requires if the magnetic entropy released across the transition is to be represented correctly. The two numbers therefore measure different physical states rather than disagreeing about the same one, and the systematic overestimate of $\beta$ by ferromagnetic saturation moments is the same effect identified in the literature curves of Figure~\ref{fig:MAP_vs_Direct}.\smallskip

Finally, we should note that Ref.~\cite{SHI2025121476} fits a Redlich-Kister expansion of $T_{\text{C}}$ for the FCC phase only and retains bare endmember values with no excess term for HCP, providing added value and novelty to our composition-dependent HCP description reported in Figure~\ref{fig:hcp_calc} and Table~\ref{tab:magnetic_params}.\smallskip

\subsection{The Revised Co-Ni Phase Diagram}
\label{sec:conidiagram}

Throughout this study, we assert that the relative stability of HCP and FCC phases in the Co-Ni binary system is particularly sensitive to magnetic entropy contributions. This assertion is validated by systematic comparisons of Co-Ni phase diagrams under various treatments of magnetic parameters, highlighting the consequences of including or excluding key magnetic descriptors in CALPHAD modeling.\smallskip

\begin{figure*}[pos=!tp]
    \centering
    \includegraphics[width=1\linewidth]{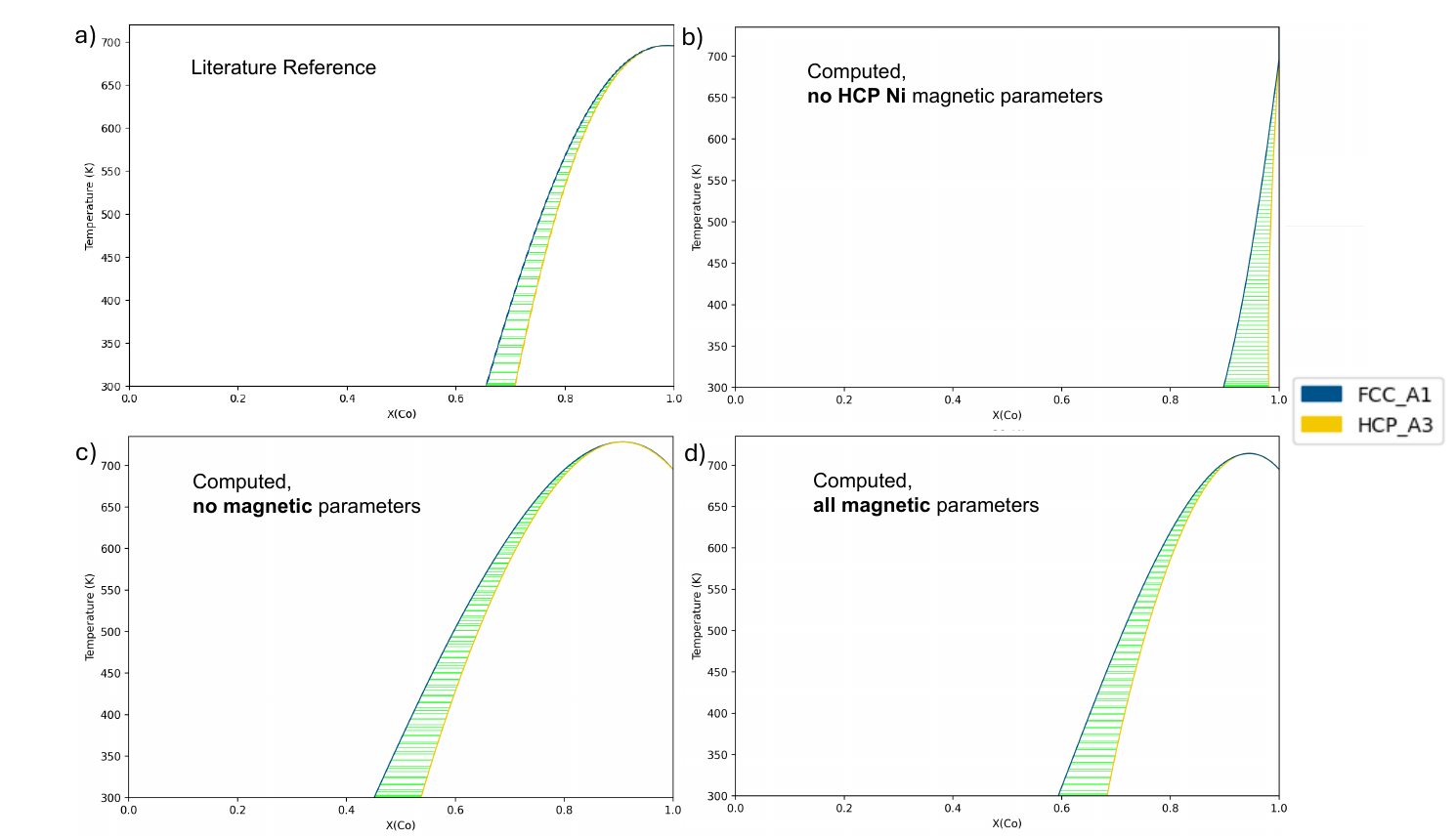}
    \caption{\textbf{a)} Co-Ni binary taken from quaternary Al-Co-Cr-Ni thermodynamic database computed by Z.K. Liu et. al. 2016~\cite{LIU2016125}, \textbf{b)} Co-Ni binary computed without HCP Ni magnetic parameters, \textbf{c)} Co-Ni binary computed with no magnetic parameters, \textbf{d)} Co-Ni binary computed with all magnetic parameters, using HCP Ni extrapolated parameters from Section \ref{sec:results}}
    \label{fig:actual}
\end{figure*}

The AlCoCrNi thermodynamic database developed by Liu et al. (2016)~\cite{LIU2016125} serves as a robust reference for benchmarking Co–Ni binary phase diagram predictions, particularly when the Co–Ni binary is extracted in isolation as shown in Figure~\ref{fig:actual}a. This database was constructed using the CALPHAD method with a combination of critically assessed experimental data and first-principles calculations, and is widely cited for high-entropy alloy (HEA) systems involving transition metals. Importantly, the Co–Ni binary interactions in this database are derived directly from the Guillermet assessment \cite{FernandezGuillermet}.\smallskip

When magnetic parameters are omitted specifically for HCP Ni, that is, when no values are defined for $\beta$ or \( T_C \) at \( X_{\mathrm{Co}} = 0 \), the calculated phase diagram exhibits a notable under-representation of the HCP region (Figure~\ref{fig:actual}b). This behavior arises because the absence of magnetic entropy contributions artificially elevates the Gibbs energy of the HCP phase in Ni-rich compositions, destabilizing it relative to FCC. As a result, the two-phase region shrinks, and the model fails to capture the correct relative stability of the HCP structure in the intermediate composition domain.\smallskip

In contrast, the diagram shown in Figure~\ref{fig:actual}c illustrates the opposite extreme in which magnetic entropy is entirely excluded from the system, i.e., no magnetic contributions are included for any phase or composition. Under this condition, the HCP phase is noticeably overrepresented. Without magnetic entropy, the stabilizing entropic contribution to the FCC phase lowers, allowing the lower-enthalpy HCP phase to dominate over a broader than expected composition range.\smallskip

When magnetic parameters are instead defined from MAP fitting of
first-principles DLM moments and Curie temperatures, the
resulting phase diagram (Figure~\ref{fig:actual}d) closely
reproduces the Guillermet derived phase boundaries in the Liu assessment
(Figure~\ref{fig:actual}a) while introducing small, physically
grounded shifts in the FCC-HCP two-phase region. The
revised HCP Ni endmembers ($T_C = 458$~K, $\beta =
0.192$~$\mu_\text{B}$) and all revised magnetic interaction parameters reduce the magnetic entropy contribution to the HCP phase relative to FCC on the Ni-rich side via the IHJ model's $R\ln(\beta + 1)$. Because the two phases no longer share identical magnetic parameters, their magnetic stabilization
diverges more with decreasing temperature: the two-phase region
widens at lower temperatures rather than maintaining the nearly
uniform width predicted by the Guillermet assessment, and shifts
slightly away from the Co-rich side, reducing overall FCC stability. The Co-rich boundary and the upper transition temperature are essentially unchanged, as expected given that the Co endmembers are fixed to SGTE values (Table~\ref{tab:magnetic_params}).\smallskip

These results demonstrate that a first-principles magnetic
description via DLM paramagnetic moments for $\beta$,
DFT-derived Curie temperatures from the corrected Curie-Weiss
model for $T_C$, independently fitted FCC and HCP interaction parameters, and DFT-determined metastable HCP Ni parameter values reproduce the experimentally validated Co-Ni phase diagram without relying on the assumptions embedded in the original assessment. The small shifts observed between Figures~\ref{fig:actual}a
and~\ref{fig:actual}d are consistent with the reduced paramagnetic
moment of itinerant ferromagnets relative to their saturation
values, and represent a more physically complete magnetic
description of the Co-Ni system that can serve as a foundation
for reassessments of higher-order systems containing Co and Ni.\smallskip

\section{Conclusion}
\label{sec:conclusion}

This work establishes a systematic first-principles methodology for constructing magnetic Gibbs energy parameters in binary systems where one or more endmembers are metastable and lack experimental magnetic data. The key methodological steps, each validated against experimental Curie temperatures and magnetic moments across the Co--Ni binary, are as follows. First, the effective magnetic moment $\beta_{\rm DLM}$ must be derived from
DLM calculations rather than FM calculations, as the DLM state directly models the paramagnetic regime that the IHJ $\beta$ parameter is meant to describe. This distinction is critical for itinerant ferromagnets like Co and Ni where the local moment is substantially reduced upon disordering. Where a constituent is non-magnetic, it enters the same effective moment of
Eq.~(\ref{eq:beta-effective}) only by dilution, with spins randomized solely
on the species carrying a finite moment. Second, the Curie-Weiss entropy expression should be used over the Spin-Counting expression, as it derives the magnetic entropy from the mean-field partition function rather than from microstate counting alone. Third, the correction $\beta_{\rm DLM}^{\rm corr} = \beta_{\rm DLM} + \exp(-\beta_{\rm DLM})$ should be applied so that low DLM moments retain a physically admissible magnetic entropy in the Curie-Weiss model. Fourth, when both endmember Curie temperatures are experimentally known, a two-point energy-based scaling provides the most accurate $T_C(x)$ across the composition range. When one or both endmember values are unavailable, the Bayesian MAP method presented here offers the next most accurate approach, anchoring the Redlich--Kister parameters to experimentally calibrated priors that carry the heat capacity information that DFT calculations cannot provide.\smallskip

Taken together, the results in Section~\ref{sec:results} underscore the necessity of
assigning magnetic parameters for metastable endmembers such as
HCP Ni and investigating mixing behavior in systems involving
such endmembers. The methodology presented here provides a
physically justified route to construct complete and internally
consistent magnetic Gibbs energy functions, ensuring that the
CALPHAD model remains predictive across complex,
composition-dependent magnetic systems.\smallskip

\appendix
\numberwithin{equation}{section}
\renewcommand{\theequation}{\thesection.\arabic{equation}}
\section{Dimensionless Piecewise Functions for the IHJ and IHX Models for Magnetic Contribution to Energy in CALPHAD}
\label{sec:appendixA}
\subsection*{$f(\tau)$ for IHJ Model}
\begin{equation}
\resizebox{\linewidth}{!}{$
f(\tau) = 
\begin{cases}
1 - \dfrac{1}{A} \left[ \dfrac{79 \tau^{-1}}{140p} + \dfrac{474}{497} \left( \dfrac{1}{p} - 1 \right) \left( \dfrac{\tau^3}{6} + \dfrac{\tau^9}{135} + \dfrac{\tau^{15}}{600} \right) \right], & \text{if } \tau < 1 \\[10pt]
-\dfrac{1}{A} \left( \dfrac{\tau^{-5}}{10} + \dfrac{\tau^{-15}}{315} + \dfrac{\tau^{-25}}{1500} \right), & \text{if } \tau \geq 1
\end{cases}
$}
\end{equation}
where:
\begin{equation*}
A = \frac{518}{1125} + \frac{11692}{15975} \left( \frac{1}{p} - 1 \right)  
\end{equation*}
The shape of $g(\tau)$ is controlled by a structure
factor $p$ (empirically $p=0.28$ for fcc/hcp and $0.4$ for bcc) which ensures continuity of $f_{\tau}$ and its derivative at $\tau = 1$, as required for a second-order transition, and partitions the magnetic enthalpy between the ordered and disordered regimes.
\subsection*{$g(\tau)$ for IHX Model}
\begin{equation}
    \resizebox{\linewidth}{!}{$
    g(\tau) = 
    \begin{cases}
    0, & \tau \leq 0 \\[10pt]
    1 - \dfrac{1}{D} \left[ 
    0.38438376 \dfrac{\tau^{-1}}{p} 
    + 0.63570895 \left( \dfrac{1}{p} - 1 \right)\left( \dfrac{\tau^3}{6} + \dfrac{\tau^9}{135} + \dfrac{\tau^{15}}{600} + \dfrac{\tau^{21}}{1617} \right)
    \right], & 0 < \tau \leq 1 \\[10pt]
    -\dfrac{1}{D} \left( \dfrac{1}{21} \tau^{-7} + \dfrac{1}{630} \tau^{-21} + \dfrac{1}{2975} \tau^{-35} + \dfrac{1}{8232} \tau^{-49} \right), & \tau > 1
    \end{cases}
    $}
\end{equation}
\begin{equation}
    D = 0.33471979 + 0.49649686 \left( \frac{1}{p} - 1 \right)
\end{equation}

\section{Explicit Derivation of \(T_C\) from DFT calculated \(E_{DLM},\) \(  E_{FM},\) \& \(\beta_{\rm DLM}\) (Curie-Weiss Model)}
\subsection{The Heisenberg Hamiltonian and Partition Function}
\label{sec:Heisenberg}
Consider the Heisenberg Hamiltonian on a $d$-dimensional lattice \cite{stanley1971introduction,yeomans1992statistical}:
\begin{equation}
\mathcal{H} = -J \sum_{\langle i,j \rangle} \vec{S}_i \cdot \vec{S}_j
\label{eq:hamiltonian2}
\end{equation}
where $J$ is the exchange coupling constant and $\langle i,j \rangle$ denotes nearest-neighbor pairs. For quantum spins with $\vec{S}_i$ being spin-$s$ operators, the canonical partition function is \cite{goldenfeld1992lectures}:
\begin{equation}
Z = \mathrm{Tr} \left[\exp\left(\frac{J}{k_B T} \sum_{\langle i,j \rangle} \vec{S}_i \cdot \vec{S}_i \right)\right]
\label{eq:quantum_partition}
\end{equation}
The Hamiltonian thus appears in the exponential as $\frac{J}{k_B T} \sum_{\langle i,j \rangle} \vec{S}_i \cdot \vec{S}_j$. Since the summation is dimensionless, the entire argument of the exponential depends only on $x = \frac{J}{k_B T}$. That is, it can be asserted that the partition functions depend only on the dimensionless parameter $x$ such that:
\begin{equation}
Z = \tilde{Z}(x), \quad x = \frac{J}{k_B T}
\label{eq:scaling_property}
\end{equation}

\subsection{Critical Temperature Scaling}
\label{sec:criticalTscaling}
A critical point occurs when the free energy or its derivatives exhibit non-analytic behavior. This manifests as loss of analyticity in the partition function. Since the partition function depends only on $x = \frac{J}{k_B T}$, any critical behavior must occur at a specific value $x_c$ of this parameter. The critical temperature $T_c$ is thus defined as the temperature where this occurs for given $J$:
\begin{equation}
x_c = \frac{J}{k_B T_c}
\label{eq:critical_parameter}
\end{equation}
Solving for $T_c$:
\begin{equation}
T_c = \frac{J}{k_B x_c} \propto J
\end{equation}
This establishes the linear scaling of $T_c \propto J$. Please note that the above derivation does not rely on any mean-field or semi-classical approximation. It only requires that the nearest-neighbor interactions dominate, so that the Hamiltonian depends on a single parameter $J$, for a given spin magnitude.\smallskip

\subsection{Ising and Curie-Weiss Hamiltonians: The Mean-Field Approximation}
\label{sec:IsingHamiltonian}
In the Ising limit (\( S_i^z = \pm \mu \), \( \mathbf{h}_i = 0 \)), the Heisenberg Hamiltonian reduces to the Ising Hamiltonian \citep{velenik2017curie}:
\begin{equation}
\mathcal{H}_{\text{Ising}} = -J \sum_{\langle i,j \rangle} S_i S_j, \quad S_i = \pm \mu
\end{equation}
The Curie-Weiss model is a mean-field approximation of the Ising model, assuming that each spin interacts equally with all others. The Hamiltonian for this model per site while correcting for double counting is \citep{velenik2017curie}:
\begin{equation}
\mathcal{H}_{\text{CW}} = -\frac{J}{2N} \sum_{i < j} S_i S_j = -\frac{Jz}{2}m^2N, \quad S_i = \pm \mu 
\end{equation}
where \(m = \langle S_i \rangle\), is mean-field magnetization or equivalently thermal average of spin and \(z\) is the coordination number of the lattice. The total mean-field energy per spin can consequently be expressed as \citep{cao2021lecture}:
\begin{equation}
\langle E \rangle = -\frac{1}{2} z J m^2 
\end{equation}

\subsection{Relating \(E_{DLM}\) and \(E_{FM}\) to thermodynamic quantities}
\label{sec:EDLMEFMthermo}
In the ferromagnetic (FM) state, an ideal perfectly aligned ferromagnet implies a system where at \(T=0\), all spins are fully aligned due to to the absence of thermal fluctuations. That is, at \(T=0\), \(S_i = +\mu \quad \forall i \implies m = \mu \). Therefore, the ferromagnetic energy (per spin) is:
\begin{equation}
    E_{FM} = -\frac{1}{2}zJ\mu^2
\end{equation}
In the high-temperature paramagnetic state approximated by the disordered-local moment (DLM) DFT results, the spins are perfectly disordered and uncorrelated, meaning that \(\langle S_i S_j \rangle = \langle S_i \rangle \langle S_j \rangle = 0 \implies m = 0\). Therefore, the paramagnetic energy (per spin) is:
\begin{equation}
    E_{DLM} = 0
\end{equation}
As a result, the exchange coupling strength J can be explicitly derived in terms of paramagnetic and ferromagnetic energy difference:
\begin{eqnarray}
    E_{DLM} - E_{FM} = 0 - (-\frac{Jz}{2}\mu^2) = \frac{Jz}{2}\mu^2\\
    \implies Jz = \frac{2(E_{DLM} - E_{FM})}{\mu^2}\\
    \therefore \quad \left(E_{DLM} - E_{FM}\right) \propto T_C \quad \because \quad J \propto T_C
\end{eqnarray}

\subsection{Deriving Scaling Factor $\frac{1}{S_{\rm mag}}$}
\label{sec:Smag_derivation}
This scaling factor is obtained by assuming a first-order magnetic transition, which is technically incorrect, but is included here for comparison. Considering Helmholtz Free Energy under fixed pressure/volume conditions, we know that:
\begin{equation*}
    F = E-TS
\end{equation*}
Considering the 2 relevant macrostates
\begin{itemize}
    \item Ferromagnetic (spin ordered) state with $E_{FM}$ and magnetic entropy $S_{FM}(T)$
    \item Paramagnetic (spin disordered DLM) state with $E_{DLM}$ and maximum magnetic entropy $S_{mag}$ associated with ideally disordered spins
\end{itemize}
At $T_C$, these 2 free energies are equal,
\begin{eqnarray*}
    F_{FM}(T_C) = F_{DLM}(T_C)\\
    E_{FM} - T_CS_{FM}(T_C) = E_{DLM}-T_CS_{\rm mag}\\
    \implies T_C = \frac{E_{DLM}-E_{FM}}{S_{\rm mag}-S_{FM}(T_C)}
\end{eqnarray*}
Under the following mean-field/ab-initio approximations:
\begin{itemize}
    \item A first order transition takes place, i.e., the ordered state near $T_C^-$ has a much smaller magnetic entropy than the maximum magnetic entropy associated with the fully spin disordered state, i.e., $S_{FM}(T_C) \ll S_{\rm mag} \implies S_{\rm mag}-S_{FM}(T_C) \approx S_{\rm mag}$
    \item Non-magnetic (vibrational, electronic) contributions to energy difference between FM and DLM states are negligible
    \item $E_{FM}$ and $E_{DLM}$ are temperature-insensitive quantities obtained from T = 0 DFT calculations that capture temperature-independent exchange coupling strength.
\end{itemize}
Therefore, we obtain that:
\begin{equation*}
    T_C \approx \frac{E_{DLM}-E_{FM}}{S_{\rm mag}}
\end{equation*}

\subsection{Relating \(\beta_{\rm DLM}\ = 2 \mu\) \text{to} \(T_C\)}
\label{sec:beta_eff_TC}
Below we re-derive the transition temperature of a Curie-Weiss model, which correctly treats the magnetic transition as second-order, under a mean-field approximation. Let \(a = \frac{Jzm}{2k_BT}\). By definition of the mean-field approximation, $m$ can be calculated as:
\begin{align}
    m &= \frac{\sum_{S_i = -\mu}^{\mu}S_i \exp{aS_i}}{\sum_{S_i = -\mu}^{\mu} \exp{aS_i}} = \frac{\partial}{\partial a} \ln(\sum_{S_i = -\mu}^{\mu} \exp{aS_i})\\
    &= \frac{\partial}{\partial a} \ln{[\exp{(-a \mu)} \sum_{S_i = 0}^{2 \mu} \exp{(aS_i)}]}\\
    &= \frac{\partial}{\partial a} \ln{[\exp{(-a \mu)} \frac{1 - e^{-(2 \mu + 1)a}}{1-e^a}]}\\
    &= \frac{\partial}{\partial a} \ln{\frac{e^{-a \mu}e^{-a/2} - e^{a \mu}e^{a/2}}{e^{-a/2} - e^{a/2}}}\\
    &= \frac{\partial}{\partial a} \ln{\frac{\sinh{(a (\mu + \frac{1}{2}))}}{\sinh{(a/2)}}}
\end{align}
Let \(a^* = \frac{Jzm}{4k_BT}\). Transforming variables,
\begin{equation}
    m = \frac{\partial}{\partial a^*} \frac{1}{2} \ln{\frac{\sinh{(a^* (2 \mu + 1))}}{\sinh{(a^*)}}} = \frac{\partial}{\partial a} f(a^*, 2\mu + 1)
\end{equation}
Solving for m consistently, the number of unique solutions (1 or 3) for the above expression depends on the value of slope w.r.t $m$ of \( f(a^*, 2\mu + 1)\) at \(a^* = 0\). Critical behavior, i.e., the magnetic transition occurs at \(T=T_C\) when \(\frac{\partial}{\partial m} f(a^*, 2\mu + 1) = 1\).\smallskip

Let $u = 2\mu + 1$ and define
\begin{equation}
    g(a^*) = \frac{1}{2}\ln\!\left(\frac{\sinh(a^* u)}{\sinh(a^*)}\right)
\end{equation}
The critical condition requires evaluating $\frac{\partial}{\partial m}g'(a^*)$ at $m=0$:
\begin{align}
    \left[\frac{\partial}{\partial m}g'(a^*)\right]_{m=0}
    &= \left[\frac{\partial a^*}{\partial m}\,g''(a^*)\right]_{m=0} \nonumber\\
    &= \left[\frac{Jz}{4k_BT}\,g''(a^*)\right]_{m=0}
\end{align}
Computing the first derivative of $g$:
\begin{equation}
    g'(x) = \frac{1}{2}\bigl[u\coth(xu) - \coth(x)\bigr]
\end{equation}
and the second derivative:
\begin{align}
    g''(x) &= -\frac{1}{2}\biggl[
        \frac{u^2}{\sinh^2(xu)}
        - \frac{1}{\sinh^2(x)}
    \biggr]
\end{align}
Expanding $\cosh^2 x = 1 + x^2 + \frac{1}{3}x^4 + O(x^6)$ about $x=0$, the ratio becomes:
\begin{align}
    &\frac{1}{2}\frac{%
        u^2(1+x^2u^2+\tfrac{1}{3}x^4u^4) - u^2}{%
        (1+x^2u^2)(1+x^2) - (1+x^2u^2)} \nonumber\\
    &\qquad{}- \frac{%
        (1+x^2u^2+\tfrac{1}{3}x^4u^4) - 1}{%
        {}- (1+x^2) + 1} \nonumber\\[6pt]
    &= \frac{3}{2}\,
        \frac{u^2 - 1}{%
            9 + 3x^2 + 3x^2u^2 + x^4u^2}
    \xrightarrow{x\to 0}
    \frac{u^2-1}{6}
\end{align}

Therefore, the self-consistency equation becomes:
\begin{equation}
    1 = \frac{Jz}{4k_BT_C}\frac{1}{6} \left(\left( 2 \mu^2 + 1\right)^2 - 1 \right)
\end{equation}
Solving for \(T_C\), we obtain the following expression, a function of J, $\mu$ (or $\beta_{\rm DLM}$), $k_B$:
\begin{align}
    T_C &= \frac{Jz}{2k_B} \frac{\mu \left(\mu + 1\right)}{3} = \frac{1}{2k_B}Jz\mu^2 \frac{\mu+1}{3\mu} \nonumber\\
        &= \frac{1}{2k_B} Jz\mu^2\frac{\beta_{\rm DLM}+2}{3\beta_{\rm DLM}}
\end{align}
Plugging in $J z \mu^2 = 2(E_{DLM} - E_{FM})$:
\begin{equation}
    T_C = \frac{E_{DLM}-E_{FM}}{k_B}\frac{\beta_{\rm DLM}+2}{3\beta_{\rm DLM}}
\end{equation}

\printcredits

\section*{Data Availability}
\label{sec:data_avail}
The thermodynamic database (TDB) file produced in this work, POSCAR files and correlation function used to generate SQS are provided in Supplementary Information. Other first-principles input and output files and the posterior sampling chains are available from the corresponding author upon reasonable request.

\section*{Acknowledgments}
\label{sec:acknowledge}
P.J. and A.v.d.W. would like to acknowledge support from NSF grants CNS-2328395 and DMR-2209027 and a Research Resilience Award from Brown University. P.J., Z.W. and J.L. would like to acknowledge funding from NASA's Aeronautics Research Mission Directorate (ARMD), Transformational Tools and Technologies (TTT) Project. We also thank Stephen Xie for engaging in useful discussions that this work benefited from.

\end{document}